\documentstyle[pre,aps,eqsecnum]{revtex}
\include{epsfig}

\tighten

\begin{document}

\draft


\title{Stochastic Hysteresis and Resonance
in a Kinetic Ising System}

\date{\today}

\author{S.~W. Sides,$^{{\rm a} \ast \dagger \ddagger}$
        P.~A. Rikvold,$^{{\rm b} \ast \dagger \ddagger}$
    and M.~A. Novotny $^{{\rm c} \dagger}$}

\address{$^{\ast}$Center for Materials Research
and Technology and Department of Physics, and \\
$^{\dagger}$Supercomputer Computations Research Institute, \\
Florida State University, Tallahassee, Florida 32306-4130 \\
$^{\ddagger}$Colorado Center for Chaos and Complexity,
University of Colorado, Boulder, Colorado 80309-0216 \\
}

\maketitle

\begin{abstract}
We study hysteresis for a two-dimensional,
spin-$1/2$, nearest-neighbor, kinetic Ising ferromagnet in an oscillating 
field, using Monte Carlo simulations and analytical theory.
Attention is focused on small systems and weak field amplitudes
at a temperature below $T_{c}$.
For these restricted parameters, the magnetization switches
through random nucleation of a {\it single} droplet of spins aligned
with the applied field. 
We analyze the stochastic hysteresis observed in this parameter regime,
using time-dependent nucleation 
theory and the theory of variable-rate Markov processes.
The theory enables us to accurately predict the results of extensive 
Monte Carlo simulations, without the use of any adjustable parameters.
The stochastic response is qualitatively different from 
what is observed, either in mean-field models or in simulations
of larger spatially extended systems.
We consider the frequency dependence of the probability density for 
the hysteresis-loop area 
and show that its average slowly crosses over to a logarithmic 
decay with frequency and amplitude for asymptotically low frequencies. 
Both the average loop area and the 
residence-time distributions for the magnetization 
show evidence of stochastic resonance. 
We also demonstrate a connection between the residence-time
distributions and the power spectral densities of the magnetization
time series. 
In addition to their significance for 
the interpretation of recent experiments 
in condensed-matter physics, including studies of 
switching in ferromagnetic and ferroelectric nanoparticles and 
ultrathin films,
our results are relevant to the general theory of periodically
driven arrays of coupled, bistable systems with stochastic noise.
\end{abstract}


\section{Introduction}
\label{sec_Intro}

Hysteretic response to an oscillating control parameter or ``force'' is a 
nonlinear nonequilibrium phenomenon commonly observed in both natural and 
man-made systems. 
The example most familiar in physics and electrical engineering 
is probably the hysteresis loop produced when a ferromagnet 
at a temperature below its critical temperature $T_c$ 
is placed in an oscillating magnetic field 
\cite{stei1892,maye91,ahai96}. 
Similar behavior is seen in ferroelectrics \cite{ishi71,orih92,beal94,rao91}.
Some other examples are electrochemical adsorbate
layers that are driven through a phase transition by an oscillating 
electrode potential in a Cyclic Voltammetry experiment 
\cite{bard80,rikv95},
systems driven through a phase transition between different 
liquid-crystalline 
phases by pressure oscillations \cite{chen96}, and systems 
driven through a solid-liquid phase transition by temperature oscillations.
Hysteresis is often modeled by systems of differential equations that display
discontinuous bifurcations \cite{nayf95,visi94,brok96}. 

Systems that exhibit hysteresis have in common a nonlinear, 
irreversible response, which causes the phase 
of the response to lag behind the force. The physical mechanism 
that causes the hysteretic behavior can, however, be quite different in 
different systems and even in different parameter regimes for the same 
system. The details of this mechanism must be considered in order to 
accurately predict such aspects of the hysteretic response as its dependence 
on the frequency and amplitude of the oscillating force. 
Here we present a study of hysteresis in a particular 
model system which incorporates both spatial degrees of freedom and 
thermal fluctuations -- a kinetic Ising ferromagnet -- in a 
parameter regime where the model has a first-order phase transition
in equilibrium
and the system response is {\it stochastic\/}. 
The model and its behavior in this regime are relevant to at least 
two different research areas that are rarely discussed together: 
experimental studies of switching dynamics in 
nanoscale ferromagnetic and ferroelectric particles and ultrathin films, 
and theoretical and experimental studies of stochastic resonance in 
spatially extended systems. We hope the present study may contribute 
to some intellectual cross-fertilization. 

In recent years 
new experimental techniques, such as magnetic force microscopy (MFM)
\cite{mart87,chan93,lede93,lede94,lede94-2}, have been developed that 
permit measurements of the magnetization state and switching behavior 
of particles as small as a few nanometers.
Ferromagnetic particles in 
this size range consist of a single domain in equilibrium, and together 
with ultrathin films they are of interest as potential 
materials for ultra-high density recording media. 
The dynamics of magnetization reversal in such systems has been 
modeled with kinetic Ising systems subject 
to sudden field reversal \cite{rich95,rich96,rich96-2,kole97,kole97-MRS}. 
These numerical and analytical 
studies have given results in qualitative agreement 
with the experiments mentioned above.
Recent experiments on ultrathin ferromagnetic Fe/Au (001)
films \cite{he93} have considered the frequency dependence of
hysteresis loop areas, which were interpreted in terms of effective
exponents consistent with those 
found for a continuous spin model \cite{rao91,rao89,rao90,rao90_2}.
Similar experiments on ultrathin Co films on Cu(001)
have found exponents consistent with a mean-field treatment
of the Ising model \cite{jiang95}. 
These studies of nanoparticles and ultrathin films
suggest that experiments can now be performed on systems 
sufficiently small that atomic-scale simulations become feasible,
and that kinetic Ising systems are useful models for switching 
in such nanoscopic systems. 

Since its introduction as a possible model for the time dependence of the 
Earth's ice ages \cite{benz81}, the concept of stochastic resonance
has been applied to a variety of phenomena in physical and biological 
science and engineering, in which 
response to a periodic force is enhanced by noise \cite{adi96}. 
Most early treatments considered a single bistable 
element similar to a mean-field model of a ferromagnet 
\cite{jung90,tome90,luse94}, with added noise. 
However, more recently experimental studies have been conducted with 
chains of coupled diode resonators \cite{loch96}, and numerical and 
theoretical studies have considered locally coupled 
one-dimensional time-dependent Ginzburg-Landau 
or Frenkel-Kontorova models \cite{benz85,lind95,lind96,marc96}, 
Ising models in one \cite{brey96}, two \cite{neda95}, 
and three \cite{neda96} dimensions, chains of coupled nonlinear maps 
\cite{pras97}, and systems of 
globally coupled bistable elements \cite{jung92}. 

Here we consider hysteresis in a two-dimensional spin-$1/2$, 
nearest-neighbor, kinetic Ising ferromagnet in an oscillating field
with periodic boundary conditions.
For convenience, and because of the many experimental measurements 
of hysteresis that address magnetic systems, we use 
the customary magnetic language, in which the order parameter 
is the magnetization per site, $m(t) \in [-1,+1]$, and the force is the 
magnetic field $H(t)$. However, we expect our results also to apply 
to stochastic hysteresis phenomena in other areas of science. 
For example, in dielectrics $m(t)$ and $H(t)$ can be 
re-interpreted as polarization and electric field, in adsorption problems 
as coverage $\theta(t) = \left[ 2m(t) - 1 \right]$ 
and (electro)chemical potential or (osmotic)pressure, etc.

Below $T_c$ and in zero field this Ising model has two degenerate
magnetized phases corresponding to a majority of the spins in the
positive or the negative direction.
A weak applied field breaks the degeneracy,
and the phase with the spins aligned (anti-aligned) with the field is 
stable (metastable). If the field varies periodically in time, the system 
is driven back and forth across a first-order phase transition, and the two 
phases alternate between being momentarily stable and metastable.
As a result, $m(t)$ lags behind $H(t)$, and hysteresis occurs.
In the regime of small system size, weak applied field, and temperature 
well below $T_c$ considered here, the system switches 
abruptly and stochastically between the two magnetized phases. 
A difference between two-dimensional, locally coupled bistable 
systems, such as this Ising model, and the one-dimensional arrays studied in 
most of the stochastic-resonance studies cited above 
\cite{loch96,benz85,lind95,lind96,marc96,brey96,pras97}, 
is that locally coupled 
one-dimensional systems have no ordered phase at nonzero temperature 
or noise intensity. The apparent long-range order in those studies is 
therefore a finite-size effect. However, the average equilibrium domain size 
grows exponentially with decreasing temperature \cite{brey96,glau63}. 
For chains much shorter than this size, the absence of true long-range 
order should not be qualitatively significant for the hysteretic behavior. 

The metastable phase in Ising
models exposed to a {\it static\/} field $H$ decays by different 
mechanisms, depending on the magnitude of $H$, 
the system size $L$, and the temperature $T$ \cite{rikv94}. 
Two distinct regimes are separated by 
a crossover field called the dynamic spinodal, $H_{\rm DSP}(T,L)$.
These two decay regimes can be distinguished by the statistical properties 
of the lifetime of the metastable phase.
The lifetime is defined as $\tau$=$t(m=0)$,
the first-passage time to a magnetization of zero,
following an instantaneous field reversal from $H$ to $-H$.
{}For $|H| \gg H_{\rm DSP}$, the mean of the lifetime, 
$\left < \tau \right >$,
is much greater than its standard deviation, $\sigma_{\tau}$.
Therefore, this field region is termed the ``deterministic regime.''
{}For $|H| \ll H_{\rm DSP}$,  
$\left < \tau \right > \approx \sigma_{\tau}$
and this field region is therefore termed the ``stochastic regime.''
Both the deterministic and stochastic regimes are further subdivided according
to the modes by which the metastable phase decays.
The deterministic regime is split into the multi-droplet (MD) and
strong-field (SF) regions for the low and high fields in this regime
respectively. For a given system size the stochastic regime
is also divided into the coexistence (CE) and single-droplet (SD)
region for the low and high fields in this regime, respectively.
Detailed discussions of these different decay modes are found
in Refs.~\cite{rikv94,rikv94_review,geilo97}.
At sufficiently low $T$ that the single-phase correlation lengths are 
microscopic, the different decay regimes 
can be distinguished by the interplay
among four length scales: the lattice spacing $a$,
the system size $L$, the radius of a
critical droplet $R_{c}$, and the average distance between
supercritical droplets $R_{0}$.
The latter two lengths increase with decreasing field strength: 
$R_c \propto 1/|H|$ and $R_0 \propto \exp[{\rm const.}/|H|^{d-1}]$, where 
$d$ is the spatial dimensionality. 
Here we consider specifically decay in the SD region, which is
characterized by
 \begin{equation}
 \label{eq_lengths}
  a \ll R_c \ll L \ll R_0 \;.
 \end{equation}
In this regime, the decay of the metastable phase proceeds by random
homogeneous nucleation of a {\it single}
critical droplet of the stable phase,
which then quickly grows to take over the system. 
We have previously proposed \cite{rich95,rich96,rich96-2,rikv94_review} 
that this decay mechanism may apply to, e.g., barium ferrite particles 
in the 50--70~nm diameter range \cite{chan93}. 
The crossover to the MD region corresponds to $R_0 \sim L$. 
As a result, the dynamic spinodal depends asymptotically on $L$ as 
$H_{\rm DSP}(T,L) \sim (\ln L)^{-1/(d-1)}$. 
In the SD region the critical droplet is much smaller than the system itself, 
and the crossover to the CE region is marked by $R_c \sim L$. 
The corresponding 
crossover field, called the thermodynamic spinodal, 
therefore depends on $L$ as 
$H_{\rm THSP}(T,L) \sim L^{-1}$. In recent exploratory 
studies \cite{side96,side97,side98-MMM} we have shown that the
response of a kinetic Ising model to an oscillating field
is qualitatively different for the MD and SD regions.
The details of the response in the MD region will be described
in forthcoming papers \cite{side98-MD}.

Theoretical studies of hysteresis
have been performed for several models, using a variety of methods.
These include various
studies of models with a single degree of freedom, equivalent to
mean-field treatments of the Ising
model \cite{jung90,tome90,luse94}, Monte Carlo (MC)
simulations of the spin-1/2 Ising model
\cite{rao90,rao90_2,acha94_review,acha92-1,acha92-2,acha94,acha94-2,lo90,fan95-2,acha97,acha97-2},
and several $O(N)$ type models \cite{rao91,rao89,rao90,rao90_2,dhar92}.
These studies were performed
with variations in the details of the simulations
and in the model parameters.
Most of them indicate that the average hysteresis-loop area, 
$\langle A \rangle $=$-\langle \oint m(H) \ dH \rangle$, appears to display
power-law dependences on the frequency and amplitude of $H(t)$. 
However, there is no universal agreement on the values of the exponents,
either experimentally or theoretically.
For the Ising model, nucleation effects that
would lead to a logarithmic frequency dependence have been
proposed \cite{beal94,rao90,thom93}.
A mean-field model exhibits a dynamic phase
transition in which the mean period-averaged magnetization,
$\langle Q \rangle $=$(\omega/2 \pi) \langle \oint m(t) \ dt \rangle$,
changes from
$\langle Q \rangle \neq0$ to $\langle Q \rangle$=$0$ \cite{tome90}.
Such a dynamic phase transition has been suggested
from MC simulations of a kinetic Ising model as well
\cite{acha94_review,acha92-1,acha92-2,acha94,acha94-2,lo90,acha97,acha97-2,acha95}.

The work presented in this paper differs from most 
past theoretical and numerical studies of hysteresis in two important ways.
First, mean-field models do not take into account
thermal noise and spatial variations
in the order parameter, thus ignoring fluctuations which may be important
in real materials. Second, most previous 
investigations of hysteresis in Ising models
have considered the frequency and amplitude dependence
of quantities such as $Q$ and $A$,
without considering the manner in which the metastable phase decays. 

Considering the nucleation-based single-droplet decay mechanism, we find that 
the average hysteresis-loop 
area exhibits an extremely slow crossover to a logarithmic 
decay with frequency and amplitude in the asymptotic low-frequency limit. 
This crossover is sufficiently slow that the behavior can easily be 
misinterpreted as a power law over several orders of magnitude in 
frequency. We also show that the average loop area and the 
residence-time distributions for the 
system magnetization exhibit evidence of stochastic resonance, and we 
provide a connection between the characteristic decay time of the 
residence-time distributions and the 
power spectral densities of the magnetization time series. 
We find no evidence of a dynamic phase transition in the SD region. 

The rest of this paper is organized as follows.
Section \ref{sec_Model} supplies background information on the simulation
of the kinetic Ising model.
In Sec.~\ref{sec_timeseries}
some general properties of the time-series data
are discussed.
In Sec.~\ref{sec_pnot}
the probability that the system magnetization
does {\it not} switch sign during
a period of the field, $P_{\rm not}(\omega)$, is derived.
This derivation is central to
theoretical calculations throughout this paper.
Section \ref{sec_residencetime} 
presents theoretical 
calculations and MC simulation data for 
the residence-time distributions (RTDs).
Also, we define and calculate
the characteristic time of the RTDs and show its relevance
to the low-frequency behavior of the power spectral densities (PSDs) of
the time series, which are analyzed in Sec.~\ref{sec_psd}.
Section \ref{sec_qa} discusses the
hysteresis-loop area, the correlation between the magnetization and the 
field, and the period-averaged magnetization. 
Finally, Sec.~\ref{sec_conclusion} contains a
summary and conclusions. 

\section{Model}
\label{sec_Model}

The model used in this study is a kinetic, nearest-neighbor
Ising ferromagnet on a
hypercubic lattice with periodic boundary conditions.
The Hamiltonian is given by
 \begin{equation}
 \label{eq_Hamil}
  {\cal H } = -J \sum_{ {\em \langle ij \rangle}} {\em s_{i}s_{j}} 
               - H(t) \sum_{i} {\em s_{i}},
 \end{equation}
where
$s_{i}$ is the state of the $i$th spin and
can have the values $s_{i}$=$\pm 1$,
$\sum_{ {\em \langle ij \rangle} }$ runs over all
nearest-neighbor pairs, and $\sum_{i}$ runs over all
$N$=$L^{d}$ lattice sites. 
The order parameter is the time-dependent magnetization per site, 
 \begin{equation}
 \label{eq_m(t)}
  {m(t)} = \frac{1}{L^{d}} \sum_{i=1}^{N} {\em s_{i}(t)} \;.
 \end{equation}

The dynamic used
is the Glauber \cite{glau63} single-spin-flip Monte
Carlo algorithm with updates at randomly chosen sites.
The time unit is one Monte Carlo step per spin (MCSS).
The system is put in contact with a heat bath at temperature
$T$, and each attempted
spin flip 
from ${\em s_{i}}$ to ${\em -s_{i}}$ is accepted with
probability \cite{bind88}
 \begin{equation}
 \label{eq_Glauber}
 W(s_{i} \rightarrow -s_{i}) = 
 \frac{ \exp(- \beta \Delta E_{i})}{1 + \exp(- \beta \Delta E_{i})} \ .
 \end{equation}
Here $\Delta E_{i}$ is the change in the energy of the system
that would result
if the spin flip were accepted,
and $\beta = 1/k_{\rm B}T$ where $k_{\rm B}$ is Boltzmann's constant. 
It has been shown in the weak-coupling limit that the 
stochastic Glauber dynamic can be
derived from a quantum-mechanical Hamiltonian
in contact with a thermal heat bath modeled as a collection
of quasi-free Fermi fields in thermal equilibrium \cite{mart77}.

In this paper, all numerical calculations
are performed for $d$=$2$, $L$=$64$ and $T$=$0.8T_{c}$. 
This value of $T$ is sufficiently far away from the critical temperature
so that the thermal correlation length is small compared to the
critical droplet radius and the size of the system. 
The system is subject to either an oscillating field, 
$H(t) = - H_0 \sin(\omega t)$, or to a constant field of magnitude $H_0$. 

As discussed in Sec.~\ref{sec_Intro}, the decay of the metastable
phase in the presence of an external field $H$ proceeds by
nucleation of droplets of the stable phase \cite{rikv94}.
Figure \ref{F(m)} shows the metastable and stable phases
as local minima in the free energy,
 \begin{equation}
 \label{eq_Fm}
  F(m,H,T) = F(m,0,T) - m L^{2} H,
 \end{equation}
of a nearest-neighbor Ising model on a $64 \times 64$
lattice at $T = 0.8T_c$ \cite{jlee95,jlee95data}. 
{}For $H$=$0$ there are two degenerate equilibrium phases of magnetization 
$\pm m_{\rm eq}(T)$, 
separated by a free-energy barrier of height proportional to $L^{d-1}$. 
{}For $H = H_0 = 0.1J$ 
the value of $m$ near $+1$, where $F$ has its global minimum,
is the stable magnetization.
The local minimum near $m$=$-1$ represents the metastable phase. 
The convex parts of the barrier represent a single spherical droplet
of one phase embedded in the other.
The droplet is a {\em collective excitation\/} \cite{jung92}
through which the switching proceeds, and the critical droplet is the 
droplet configuration corresponding to the local maximum of $F$ 
at a given value of $H$ \cite{jlee95}. 
For $H<0$ the stable and metastable phases are reversed. 

The average number of droplets of the stable phase
that are formed per unit time and volume is given by the
field and temperature dependent nucleation rate,
 \begin{equation}
 \label{eq_I}
  I \left ( H(t),T \right ) \approx B(T) |H(t)|^K 
                   \exp \left [- \frac{\Xi_{0}(T)}{|H(t)|^{d-1}} \right ].
 \end{equation}
The notation follows that of Ref.~\cite{rich95}, where $B(T)$ is a 
non-universal temperature dependent prefactor, and $K$ and $\Xi_{0}(T)$
are known from field theory \cite{lang67,lang69,gnw80}
and simulations \cite{rikv94} and are
listed in Table \ref{table_constants}.
The quantity $\Xi_{0}(T)$ is the field-independent part of the
free-energy cost of a
critical droplet, divided by $k_{B}T$.
The external field, $H(t)$, is the only quantity through which
$I \left ( H(t),T \right )$ depends on time in this adiabatic approximation.

Several quantities, whose values do not depend on the frequency
of the applied field, are required as input for the theoretical
calculations in the following sections.
These quantities, which include the average lifetime 
$\left < \tau (H_0) \right > \approx [L^d I(H_0,T)]^{-1}$, 
are listed in Table \ref{table_constants}.
They are determined through what we refer to as
``field-reversal simulations.''
In these simulations the system initially has all spins up or positive.
It is then subjected to a static external field of magnitude $H_0$ 
with a sign opposite the system magnetization. This instantaneous
field quench prepares the system in a metastable state, and the
decay of the metastable phase proceeds by the mechanisms outlined in
the introduction.

\section{Time-Series Data}
\label{sec_timeseries}

\begin{figure}[tbp]
\vspace*{3.5in}
\includegraphics{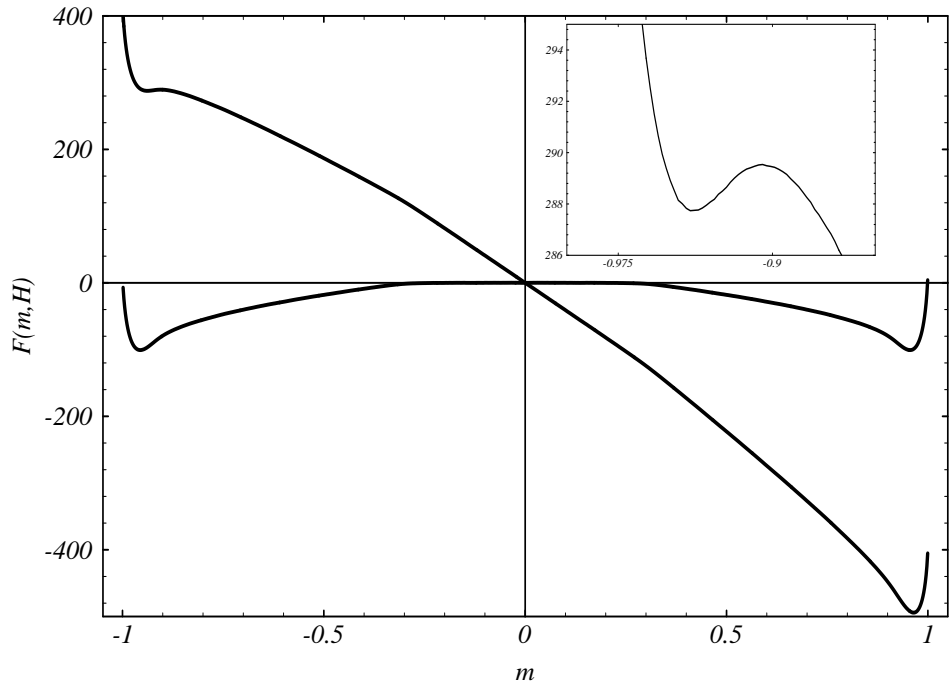}
\bigskip
\caption{ \label{F(m)}
The free energy, $F(m,H,T)$, shown vs.\ magnetization $m$
for the nearest-neighbor
Ising ferromagnet on a $64 \times 64$ square lattice at $T$=$0.8T_{c}$.
Data are shown for $H$=$0.0J$ and $0.1J$.
The data was obtained from a study in which
a multicanonical MC algorithm was used to find $F(m,0,T)$
\protect\cite{jlee95}.
The inset shows an expanded view of the portion of the free energy curve
near the metastable state for $H$=$0.1J$.
The barrier height is on the order of $1 k_{\rm B} T$.
}
\end{figure}

In the simulations presented here,
a sinusoidal field is applied to the system.
Its amplitude, $H_{0} = 0.1J < H_{\rm DSP}$, is chosen
such that in field-reversal simulations the system is clearly
in the SD region for a field of magnitude $H_{0}$.
The dynamic spinodal field is approximated by $H_{\rm DSP} \approx H_{1/2}$,
where $H_{1/2}$ is the value of $H$ (for given $L$ and $T$) for which
the relative standard deviation of the lifetime,
$r$=$\sigma_{\tau}/\left < \tau \right >$, is $1/2$ \cite{rikv94}.
This value of $H_{\rm DSP}$ is given in Table \ref{table_constants}.	
It approximately equals the field for which the local minimum
in Fig.~\ref{F(m)} disappears \cite{jlee95}.

To obtain the raw time-series data, an Ising system was
initially prepared with either a
random arrangement of up and down spins with $m(t=0) \approx 0$,
or with a uniform arrangement with all spins up.
Then the sinusoidal field, $H(t)$=$-H_{0} \sin(\omega t)$, was applied
and changed every attempted
spin flip, allowing
for a smooth variation of the field.
The time series did not appear to depend on the initial conditions
after a few periods.
The simulations were performed
with several values of the driving frequency $\omega$.
For each frequency, we recorded the time-dependent magnetization $m(t)$ 
for approximately $16.9 \times 10^{6}$ MCSS. 
Each of these raw time-series data files
store the values of $t$, $H(t)$, and $m(t)$ in increments of $1$ MCSS.
Each file takes up about $800$ megabytes and took about $9$ days to run,
using a single node of an IBM sp2. 
These are by far the most extensive MC simulations of hysteresis 
in Ising systems to date. 

It is useful to think of hysteresis as a competition between two time
scales: the average lifetime of the metastable state
following an instantaneous field reversal from $H_{0}$ to $-H_{0}$,
$\left < \tau(H_{0}) \right >$, and the period of
the external forcing field, $2 \pi / \omega$.
Therefore, we specify the ratio $R$ of the period to the average lifetime,
 \begin{equation}
 \label{eq_R}
   R = \frac{(2 \pi / \omega)}{\left < \tau(H_{0}) \right >} \ .
 \end{equation}
One may think of $R$ ($1/R$) as a scaled period (scaled frequency). 

We note that $\langle \tau(H_0) \rangle$ 
is the ``shortest of the long time scales'' in the 
present system.
From Fig.~\ref{F(m)} we observe that whereas the free-energy 
barrier represented by the critical droplet for $|H|=H_0$ is on the order of 
$k_B T$, the barrier between the degenerate phases at
$H=0$ is on the order of $50 k_B T$.
For this temperature and system size,
the timescale for spontaneous fluctuations
between the phases
in the absence of an applied field, $\left < \tau(0) \right >$,
is therefore essentially infinite.
Conversely, the nucleation of the critical 
droplet necessary to leave the metastable phase is entirely driven 
by the thermal fluctuations, even when the field has its 
maximum strength $H_0$.
Switching in this system 
therefore truly depends on the joint action of the random thermal 
noise and the deterministic oscillating field. 

Figure \ref{mtSD} shows short initial segments of the magnetization
time series in the SD region for three different values of $R$.
In all three cases, $m(t)$ fluctuates near one
of the two degenerate values of the spontaneous zero-field magnetization,
punctuated by rapid transitions between these two values,
that are completed during a {\it single} half-period
of the applied field.
The rapid switching
of $m(t)$ is evidence of the nucleation of a single critical
droplet that reverses the sign of the magnetization.
The magnetization `plateaus' are due to the failure of any critical
droplet of the stable phase to nucleate.
The fluctuations in $m(t)$ on these plateaus indicate
appearance and disappearance of sub-critical droplets. 
For low driving frequencies the magnetization switches twice during almost 
every field cycle, whereas for high frequencies switching occurs only 
occasionally. 
The `spikes' in $m(t)$ seen for $R$=$2.5$
occur when a single droplet nucleates,
but does not have time to grow and switch the system
magnetization before the field becomes unfavorable
and the droplet rapidly collapses.
The number of field cycles shown in Fig.~\ref{mtSD}
is small compared to the total number of cycles in an
entire time series.

\section{Probability of not switching during a period}
\label{sec_pnot}

The probability that the system does not switch during a full period
of the field, $P_{\rm not}(\omega)$,
is central to the theoretical understanding of
hysteresis in the SD region.
It occurs most directly in the calculation of the
residence-time distributions in Sec.~\ref{sec_residencetime}.
In addition, elements in the derivation of
$P_{\rm not}(\omega)$ are fundamental for
describing most of the observed quantities in this study.

As mentioned in Sec.~\ref{sec_timeseries}, 
the system exhibits
abrupt switches during which the average magnetization
changes between values near $\pm m_{\rm eq}$.
As seen in Fig.~\ref{mtSD},
these events occur quickly compared
to the period of the external field,
and to a first approximation
the time it takes the droplet to grow to fill the system (the growth time)
is negligible.
A more realistic treatment takes the finite growth time into account
as a lag time between the nucleation of a critical droplet
and the time at which the system switches.

The first part of the derivation is presented without the
effects of the growth time.
This is done for simplicity, as well as to
emphasize the role of the growth time as a {\it correction}
to the basic picture of a
variable-rate Poisson process.
First we derive the expression for the cumulative probability
that a switching event has occurred by time $t$, $F(t)$,
in terms of the time-dependent rate
of an {\it instantaneous} decay process, $\rho(t)$. 
[This cumulative probability 
should not be confused with the free energy $F(m,H,T)$ of Eq.~(\ref{eq_Fm}).] 
It is convenient to introduce $\bar{F}(t) = 1 - F(t)$, 
the probability that a switching event
{\it has not} occurred by time $t$.
Standard theory of variable-rate Markov processes \cite{cox65}
leads to a difference equation for $\bar{F}(t)$,
 \begin{eqnarray}
   \nonumber
   \bar{F}(t+\Delta t) & = & \bar{F}(t) \times \lbrace 
                             \mbox{probability an event has not 
                   occurred in the interval}\,[t,t+\Delta t] \rbrace  \\
                      \label{eq_F(t)nogrowth2}
                 & = & \bar{F}(t) \left [ 1- \rho(t) \Delta t \right ] \ ,
 \end{eqnarray}
which in the limit $\Delta t \rightarrow 0$ gives
 \begin{equation}
  d\bar{F}(t)/dt = -\rho(t) \bar{F}(t).
 \end{equation}

The growth time, $t_{\rm g}(t)$, is introduced into the derivation
at the level of Eq.~(\ref{eq_F(t)nogrowth2}).
It is defined as the
time between the nucleation of a critical droplet
and the time when the volume of this droplet
becomes approximately half the system volume.
The dependence of the
growth time on $t$ is a consequence of the time dependence
of the interface growth velocity, which is approximately
proportional to $H(t)$.
For suitably long time scales, the growth of a supercritical
droplet is a deterministic process.
Another  quantity in this derivation
is the time at which a droplet nucleates, $t_{\rm n}(t)$.
If a switching event occurs at time $t$, then
$t_{\rm n}(t)$=$t - t_{\rm g}(t)$.
Where clarity is not sacrificed, we do not show the explicit
$t$ dependence of $t_{\rm g}$ or $t_{\rm n}$.
For a switching event to occur in any particular
period of the external field, a critical droplet must not only nucleate,
but must do so early enough so there is sufficient time for it
to grow to the volume of the system.
Therefore, the difference equation for ${\bar F}(t)$ is
modified to read
 \begin{eqnarray}
  \nonumber
   \bar{F}(t+\Delta t)
      & = & \bar{F}(t) \times
      \lbrace \mbox{probability
                    a switch has not occurred within}\,[t,t+\Delta t]
                    \rbrace \\
    \nonumber
      & = & \bar{F}(t) \times
      \lbrace \mbox{probability
                    a droplet has not nucleated within}\,
                    [t_{\rm n},t_{\rm n}+\Delta t_{\rm n}] \rbrace \\
     \nonumber
          & = & \bar{F}(t) \left [ 1- \rho(t_{\rm n})
                \Delta t_{\rm n} \right ], \\
  \label{eq_F(t)growth1d}
          & = & \bar{F}(t) \left [ 1- \rho(t_{\rm n}) \frac{dt_{\rm n}}{dt} 
                \Delta t \right ].
  \end{eqnarray}
We can express this result in terms of the growth time $t_{\rm g}$
and its derivative, using
 \begin{equation}
 \label{eq_derivativegrowthtime}
 \frac{dt_{\rm n}}{dt} = 1 - \frac{dt_{\rm g}}{dt} \ .
 \end{equation}
Substituting into Eq.~(\ref{eq_F(t)growth1d}) and letting
$\Delta t \rightarrow 0$ gives
 \begin{equation}
 \label{eq_F(t)growth1_last}
  \frac{d\bar{F}(t)}{dt} = -\rho(t-t_{\rm g}) 
       \left[1- \frac{dt_{\rm g}}{dt} \right ] \bar{F}(t).
 \end{equation}
Integrating Eq.~(\ref{eq_F(t)growth1_last}) gives the cumulative
distribution,
  \begin{equation}
  \label{eq_F(t)growth2}		
    F(t)  =  1 - \exp \left [- \int_{0}^{t} \rho[t'-t_{\rm g}(t')] 
               \Bigl(1-\frac{dt_{\rm g}(t')}{dt'} \Bigr ) 
               dt' \right ].
  \end{equation}
Differentiation gives the probability density function (pdf) for
switching events at time $t$,
 \begin{equation}
 \label{eq_F(t)growth3}
  P(t) = 
  \frac{dF}{dt} 
        = \rho [t-t_{\rm g}(t)] \left [1-\frac{dt_{\rm g}(t)}{dt} \right ]
           \exp \Biggl [-\int_{0}^{t} \rho[t'-t_{\rm g}(t')] 
           \biggl(1-\frac{dt_{\rm g}(t')}{dt'} 
            \biggr) dt'
           \Biggr ].
 \end{equation}
For $t_{\rm g}$=$0$, Eq.~(\ref{eq_F(t)growth3}) is equivalent
to Eq.~(9) of \cite{zhou90-2}.
The growth time $t_{\rm g}$ is obtained from the expression for
the time-dependent volume of a supercritical droplet
 \begin{equation}
 \label{eq_V(t)1}
  V(t,t_{\rm n}) = \Omega \left [ \int^{t}_{t_{\rm n}} v(t') dt' 
                    \right ]^{d} ,
 \end{equation}
where $t > t_{\rm n}$.
Here $v(t)$ is the droplet interface velocity and
$\Omega$ is defined such that the volume of an equilibrium droplet of radius
$R$ is $\Omega R^{d}$ \cite{ccga94}.
Using the
Lifshitz-Allen-Cahn approximation \cite{gunt83,lifs62,alle79,seki86},
the interface velocity is $v(t) \approx \nu |H(t)|$.
The proportionality constant $\nu$ depends on the details
of the dynamics. Here we use values for the Glauber dynamics, obtained
from field-reversal simulations
by Ramos {\it et al.\/} \cite{ramo97}.
The values of the constants used in our calculations are
listed in Table \ref{table_constants}.
For $d$=$2$, Eq.~(\ref{eq_V(t)1}) for the growth time becomes,
 \begin{eqnarray}
  \nonumber
   V(t,t-t_{\rm g}) = \frac{L^{2}}{2}
     & = & \Omega \left [ \int^{t}_{t-t_{\rm g}} 
     \nu H_{0} \sin \omega t' dt' \right ]^2 \\
     \label{eq_V(t)2}
     & = & \frac{\Omega \nu^2 H_{0}^{2}}{\omega^2} \Bigl \{ \cos \omega 
                [t-t_{\rm g}] - \cos \omega t \Bigr \}^2.
 \end{eqnarray}
\noindent
For a static field of strength $H_{0}$, the growth time is 
 \begin{equation}
 \label{eq_static_tg}
  \tilde{t}_{\rm g} 
= \frac{1}{\sqrt{2 \Omega}} \left ( \frac{L}{\nu H_{0}} \right ).
 \end{equation}
Substituting this expression into Eq.~(\ref{eq_V(t)2}) gives
 \begin{equation}
  \cos \omega [t-t_{\rm g}] = \tilde{t}_{\rm g} \omega + \cos \omega t.
 \end{equation}
Solving for $t_{\rm g}$ such that $t_{\rm g} < t$ gives
  \begin{equation}
  \label{eq_tg}
  t_{\rm g}(t)  =  \left \{ \begin{array}{ll}
               t - \{ \frac{1}{\omega} \cos^{-1} 
                 [ \cos \omega t + \tilde{t}_{\rm g} \omega] \} &
                 \ \ \ \ \ t_{0} < t < \pi / \omega \\
               0 & \mbox{\ \ \ \ \ otherwise}
                        \end{array},
               \right.
  \end{equation}
where 
 \begin{equation}
 \label{eq_to}
  t_{0} = \frac{1}{\omega} \cos^{-1} [1- \tilde{t}_{\rm g} \omega].
 \end{equation}
The time $t_{0}$ is the first time during a period for which the
probability of switching is non-zero.
If $H(t$=$0)$=$0$ and $m(t$=$0) \approx -1$, then the
probability density $P(t)$
that a switching event takes place at time $t$ is
  \begin{equation}
  \label{eq_P(t)}
   P(t) = \left \{ \begin{array}{ll}
      0 & 0 < t < t_{0} \\
      \rho [t-t_{\rm g}(t)] \Bigl [ 1-\frac{dt_{\rm g}(t)}{dt} \Bigr ]
      \exp \Bigl (-\int_{0}^{t} \rho[t'-t_{\rm g}(t')] 
      \left [1-\frac{dt_{\rm g}(t')}{dt'} \right ]
            dt' \Bigr ) &
                t_{0} < t < \pi / \omega \\
       0 & \mbox{$\pi / \omega < t < 2 \pi / \omega$,}
    \end{array}
          \right. 
  \end{equation}
where 
the ranges for $t$ ensure $P(t)$ is non-zero only when
$t > t_{\rm g}$ and the signs  of $m(t)$ and $H(t)$ are not equal.
Higher-order corrections, including the probability that a second
droplet nucleates during $t_{\rm g}$, were found to be numerically
insignificant.
The main approximation used here to obtain $t_{g}(t)$ lies
in ignoring the slower growth of droplets only slightly
larger than the critical radius.
This has been shown to be permissible for adiabatically slow-forcing
models in which large droplets grow exponentially in time
\cite{shne94-2}.
In the present case, however, we simply consider it a convenient
approximation, whose accuracy is ultimately confirmed by our
numerical simulations.

In the SD region, the average lifetime in a field-reversal simulation
should be dominated by nucleation.
Therefore, the total nucleation rate
in a static field $H_{0}$ can be expressed as
 \begin{mathletters}
 \begin{eqnarray}
 \label{eq_rho_tau}
  \rho_{0}
         & = & \left [ \left < \tau \right > - \tilde{t}_{\rm g} 
               \right]^{-1} \\
 \label{eq_rho_tau_b}
         & = & L^{d} I \left ( H_{0},T \right ).
 \end{eqnarray}
 \end{mathletters}
The nucleation rate in a static field of strength $H_{0}$ should be
equal to the nucleation rate in a sinusoidal field of 
amplitude $H_{0}$ at the maximum
of the field,
$\rho_{0}$=$\rho(t=\pi/2 \omega)$.
The ratio of these two decay rates is then
 \label{eq_rate_ratio}
  \begin{equation}
   \frac{\rho(t)}{\rho_{0}} = 
   \frac{ I \left(H(t),T \right )}{ I \left(H_{0},T \right )}.
  \end{equation}
Substituting the form of the nucleation rate, Eq.~(\ref{eq_I}),
into the expression above
allows $\rho(t)$ to be recast in a form which does not explicitly
contain the non-universal prefactor $B(T)$:
 \begin{equation}
 \label{eq_rho(t)}
 \rho(t) = \rho_{0}
           |\sin (\omega t)|^K \exp
           \left [ - \frac{\Xi_{0}(T)}{|H_{0}|^{d-1}} 
           \left ( \frac{1}{|\sin (\omega t)|^{d-1}} -1 \right ) \right ] \ .
 \end{equation}
This expression holds when $m(t)$ and $H(t)$ have opposite signs,
while $\rho(t)$=$0$ when they have the same sign.
Using Eq.~(\ref{eq_rho_tau}) for the maximum decay rate gives
$\rho_{0}$=$(6.62 \pm 0.07) \times 10^{-4} \ {\rm MCSS}^{-1}$,
using quantities listed in Table \ref{table_constants}.
Figure \ref{fig_rho}
shows $P(t)/ \omega$ vs.\ $\omega t$ for five different frequencies of the
external field.
The inset in Fig. \ref{fig_rho} shows $\rho(t)$ vs.\ $\omega t$.
The nucleation rate achieves its maximum value, $\rho_{0}$,
at a phase of $\omega t$=$\pi/2$, independent of $\omega$.
However, the location and width of the maximum for $P(t)$ depend
strongly on $\omega$.
This behavior results from
the combined field dependence of the nucleation rate
and the interface growth velocity.
For $R \gtrsim 100$, $P(t)$ narrows and the location of
its maximum shifts to lower phase values as the
switching begins to occur before the maximum in $\rho(t)$.
As $R$ is decreased below $1.5$,
$\omega t_{0} \rightarrow \pi$, and the area under
the curve for $P(t)/\omega$ goes to zero.
Therefore, the condition $\omega_{\rm max} t_{0}$=$\pi$
gives the maximum frequency for which single-droplet
switching is possible.
Using Eq.~(\ref{eq_to}) and converting the result to a
bound on $1/R$
gives $(1/R)_{\rm max}$=$1.19 \pm 0.03$.
For higher frequencies, switching events are very rare,
and if they occur at all, they do so through a multidroplet mechanism.
The probability of not switching during an entire period is
obtained by integrating the probability density $P(t)$,
 \begin{eqnarray}
 \nonumber
  P_{\rm not}(\omega) 
     & = & 1 - \int_{0}^{\frac{\pi}{\omega}} P(t') dt' \\
    \label{eq_Pnot}
     & = & \bar{F} \left( \frac{\pi}{\omega} \right ) \ .
 \end{eqnarray}
The frequency dependence of $P_{\rm not}(\omega)$
is the aspect of this quantity 
most important for comparisons with our MC data.
However, $P_{\rm not}(\omega)$ also
depends on other parameters through the nucleation rate.

\section{Residence-time analysis}
\label{sec_residencetime}

As mentioned in Sec.~\ref{sec_timeseries}, the magnetization
exhibits abrupt switches between values near $\pm m_{\rm eq}$.
In contrast, the times between the magnetization
reversals are comparable to, or greater than, the
period of the applied field.
As a first approximation, one may therefore consider
these switching events as occurring in a discrete
two-state system.
For an Ising system
undergoing a field-reversal experiment \cite{rikv94},
the lifetime in the SD region is stochastic and
is well described by droplet theory.
For an oscillating field, the analogous quantity
is the time between reversals of the magnetization,
called the {\it residence time}.
The probability density for the residence times is called
the residence-time distribution (RTD) \cite{jung93,gamm95}.
In Sec.~\ref{subsec_residencetime_dis}
we construct analytical expressions for
the RTDs and compare these with the RTDs obtained from our
simulated time series.
In Sec.~\ref{subsec_residencetime_strength} we calculate the area of
the peaks in the RTDs, or the RTD peak strengths,
and compare our theoretical results for the peak strengths
with MC data.
Finally, we show that our data for the RTD peak strengths provide
evidence of stochastic resonance in the model.

\subsection{Residence-time distributions}
\label{subsec_residencetime_dis}

We define the residence time, $\Delta$, as the time between
consecutive magnetization reversals,
and denote its probability density as $\Pi(\Delta)$.
The details of our theoretical deviation of $\Pi(\Delta)$
are given in Appendix \ref{appendix_residencetime_dis}.
The results of the theoretical calculation for the residence-time
distributions, which contain \underline{no} adjustable parameters,
are shown as solid curves
in Fig.~\ref{rtdSD}(a-c) for different values of $R$.

Next, we give a description of the MC analysis for the RTDs.
(The results of this analysis are shown as solid dots
in Fig.~\ref{rtdSD}(a-c) for different values of $R$.)
When measuring an RTD one must ignore ``false crossing events.''
In these events, the magnetization crosses
zero and re-crosses zero again within a short time
without having reached a value near the stable magnetization.
There appear to be two reasons for these re-crossing events.
First, when $m(t) \approx 0$ the magnetization can re-cross zero many times
due to thermal fluctuations.
Second, during some of the periods there will
be a ``spike'' in the magnetization when a supercritical droplet
nucleates but does not have time to completely
take over the system before the applied field changes sign.
For this reason, a cutoff is employed, and a switching event
is recorded only
when $m(t)$ reaches some cutoff value
$\pm m_{\rm cut}$.
To quantify the meaning of the cutoff, define $t^{+}_{i}$ and $t^{-}_{i}$ 
as the times at which
$m(t^{+}_{i})$=$+m_{\rm cut}$ and
$m(t^{-}_{i})$=$-m_{\rm cut}$, respectively.
The residence time $\Delta_{i}$ is given by
 \begin{equation}
   \Delta_{i} = \left \{ \begin{array}{ll}
   t^{+}_{i+1} - t^{-}_{i} \ \ \ {\mbox {\rm when}} \ m(t) \approx -1 \ \ 
                                             {\mbox {\rm for}} \ \
                                         t^{-}_{i} < t < t^{+}_{i+1} \\
   t^{-}_{i+1} - t^{+}_{i} \ \ \ {\mbox {\rm when}} \ m(t) \approx +1 \ \ 
                                             {\mbox {\rm for}} \ \
                                         t^{+}_{i} < t < t^{-}_{i+1}
                     \end{array}.
               \right.
 \end{equation}
We used $m_{\rm cut}$=$0.25$.
For each frequency, the residence times are measured over
an entire time series.
The size of the bins in an RTD is set by dividing the
maximum observed residence time by the number of bins.
Both the maximum residence time for a given time series
and the number of bins are
different for different frequencies.
Hence, the size of the bins is different for each of
the graphs in Fig.~\ref{rtdSD}.
Scaling the residence times 
by the period of the external field centers the peaks
in the RTD about every odd half-integer.
In the low-frequency limit, the system
spends enough time in an unfavorable field
during every half-period to allow the magnetization
to switch.
In this limit, the RTD would contain a
single peak centered around $1/2$.
As the frequency of the field is increased, there
should be more periods during which the magnetization does not
switch at all, indicated in the RTDs by an increase of the size
of the peaks centered on $3/2$, \  $5/2$, \  etc.
The RTD data from MC simulations are shown in Fig.~\ref{rtdSD}
as solid points together with the theoretical curves.

The smoothest MC results, {\it and} the best agreement between
theory and simulation occur for $R \agt 10$.
The agreement is quite good, considering that the theoretical
calculation contains {\it no} free parameters.
All of the constants in the formulas for the nucleation rate and the
interface growth velocity come from theoretical considerations or
from field-reversal simulations.
The agreement is poor only for the highest
frequencies, corresponding to $R \lesssim 2.5$.
For these values of $R$ the MC data are suspect.
First, for these high frequencies of the external field, the sizes
of the peaks for large residence times are significant.
For all of the frequencies shown, the data sets have approximately
the same total number of switching events.
Therefore, a smaller number of events is contained in each bin
of the RTDs for the higher frequencies.
Second, in spite of the cutoff there are two peaks in the RTDs
for $R$=$2$ through $5$ in the interval $0 < \omega \Delta / 2 \pi < 1$.
Of these two peaks, the peak at the shorter residence time
comes from ``spikes'' in the magnetization which are large enough
to extend past the cutoff value, but still do not switch the system
completely within a field period.
These ``spikes'' in the time series
redistribute weight from the higher-order peaks of the RTD into
the spurious peak at short residence times.
This affects the measurement of the peak strengths from the MC data as well,
as discussed in the next section.

\subsection{RTD peak strengths}
\label{subsec_residencetime_strength}

The frequency dependence of the strengths of the peaks in the RTD is
another quantity which describes
the nature of the magnetization reversal.
The strength of the $j$-th peak in an RTD is given by
 \begin{equation}
 \label{eq_rtdpeak_def}
  S_{j}(\omega)
  = \int_{(j-1)\frac{2 \pi}{\omega}}^{j \frac{2 \pi}{\omega}} 
    \Pi (\Delta) d \Delta.
 \end{equation}
The peak strengths obtained 
from the MC time series are shown as solid dots in Fig.~\ref{rtdSDpeak}.
The statistical errors are estimated by error-propagation analysis
as $\sqrt{S_{j} (1-S_{j})/ N}$, where $N$ is the total
number of switching events in a simulation run.
For almost all of the data points the error bars are smaller
than the symbol size. The solid lines
in Fig.~\ref{rtdSDpeak} are theoretical results.
The strength of the first peak, $S_{1}(\omega)$,
is simply the probability that $m(t)$ switches sign {\it within}
the first period after the last field reversal,
 \begin{equation}
 \label{eq_S1}
  S_{1}(\omega) = 1 - P_{\rm not}(\omega) \ .
 \end{equation}
Therefore, the strength of the $j$-th peak is
 \begin{equation}
 \label{eq_Somega}
   S_{j}(\omega) = P_{\rm not}(\omega)^{j-1} 
   \left [ 1 - P_{\rm not}(\omega) \right ].
 \end{equation}
The values for $P_{\rm not}(\omega)$ used in this calculation
were obtained by numerically integrating Eq.~(\ref{eq_Pnot}) for several
values of $\omega$.
This parameter-free numerical evaluation of the theoretical peak strengths 
is in good agreement with the MC data. 
In Fig.~\ref{rtdSDpeak} one can see this agreement, especially
for the strength in the first peak, $S_{1}(\omega)$, for all
but the highest-frequency data point at $1/R$=$0.5$.
However, the MC data slightly overestimate $S_{1}(\omega)$
even for low frequencies.
This is due to the redistribution of strength from the higher-order
peaks into the first peak due to ``spikes'' in the time series
which extend past the cutoff, as mentioned in
Sec.~\ref{subsec_residencetime_dis}.
Hence, the peak strengths for the higher-order peaks
are systematically underestimated by the MC data, particularly
for $S_{2}(\omega)$.
The agreement between the theoretical curve and the data
is not quite as good for $S_{2}(\omega)$, $S_{3}(\omega)$, and
$S_{4}(\omega)$ at higher frequencies.
However this is expected, due to the
poorer statistics for these higher-order peaks.

Analysis of the RTDs for two-state systems has been used to detect
stochastic resonance (SR) \cite{adi96,jung93}.
Gammaitoni {\it et al.\/} studied the switching behavior and residence
times of an analog circuit, which served as a model 
of a bistable system driven by random noise and a 
sinusoidal external forcing \cite{gamm95}.
They found for their model that SR is manifest in the fact that each of
the peak strengths in the RTDs
has a maximum for a given frequency of the field.
If the frequency $\omega_{j}$ corresponds to the location of the maximum
in $S_{j}$, then
 \begin{equation}
 \label{eq_omegas}
  \omega_{1} < \omega_{2} < \omega_{3} < \ldots \; .
 \end{equation}
\noindent
This approach gives an alternative definition for SR, 
different from the original definition 
as the noise intensity for which the signal-to-noise ratio (SNR)
exhibits a maximum \cite{adi96}.
Since the SNR does not exhibit a maximum with respect to the
frequency of the forcing, the definition suggested by Gammaitoni 
{\it et al.\/} facilitates the understanding of SR as the tuning of 
one timescale (inverse frequency)
to another (average lifetime of the metastable phase), more 
analogous to a ``bona fide'' resonance. 

The case studied in Ref.~\cite{gamm95} was one of a very weak oscillating 
field, such that the noise-driven switching rate in zero field 
was of the same order of 
magnitude as the escape rate from the metastable state in maximum field. 
As a result, there was only {\em one\/} long timescale, 
and for sufficiently low frequencies most of the escapes were completely 
thermally driven. This would redistribute the number of escapes with 
residence times less than $\pi/\omega$ into a peak at much shorter times, 
corresponding to the thermal escape rate of the unforced bistable system 
and giving a nonzero value of $\omega_1$. 

For the system studied here
the thermal switching rate in zero field is virtually zero, 
as pointed out in the discussion following Eq.~(\ref{eq_R}). 
As a result, $S_{1}$ increases monotonically towards unity 
as the driving frequency is lowered.
For $S_{1}$ to display a maximum, the weight in the RTD's
must shift towards times much shorter than the period
of $H(t)$.
These residence times would correspond to events in which
a single thermal fluctuation switches the system.
One can increase the thermal switching rate in zero field
(or decrease the thermal relaxation time in zero field
$\langle \tau(0) \rangle$)
by increasing $T$.
(Recent measurements of telegraph noise in nanoscale ferromagnetic
particles are able to show this quite clearly \cite{wern97-1}.)
But increasing the temperature can move the
system into an entirely different
decay regime (either the MD or SF regime, depending on the
value of $T$) where the stochastic nature of the response disappears.
However, even though $S_{1}$ does not display a maximum,
the higher-order peak strengths do have maxima at 
$\omega_{2} < \omega_{3} < \ldots$.
From Eq.~(\ref{eq_Somega})
the theoretical positions of these maxima are seen to be given by the 
condition $P_{\rm not}(\omega_j) = (j-1)/j$, which yields $\omega_1 = 0$. 
Through $P_{\rm not}(\omega)$ they are determined by the 
competition between the period of the deterministic forcing and a 
stochastic timescale, which is the 
{\it minimum\/} metastable lifetime $\langle \tau(H_0) \rangle$.

If we were to reduce $L$ 
to push the simulations into the coexistence (CE) region,
then $\langle \tau(0) \rangle$,
which depends exponentially on $L^{d-1}$ through the barrier in the 
free energy $F(m,0,T)$, would approach 
the escape time $\langle \tau(H_0) \rangle$, which 
by Eq.~(\ref{eq_rho_tau_b}) is 
inversely proportional to $L^d$.
As a result, 
$S_{1}$ should be observed to decrease at very low frequencies for 
sufficiently small systems.
In this regime, the critical droplet volume 
would be on the order of half the system volume.
This effect was 
recently observed by Lindner {\it et al.\/} in simulations of a 
one-dimensional chain of bistable elements driven at a constant 
frequency and subject to noise of variable intensity \cite{lind95,lind96}. 

Our results lead us to make the observation that a {\em maximum\/} in 
$S_1$ is not necessary for the response of the system to the oscillating 
field to be characterized as ``resonant.'' 
Rather, $m(t)$ is essentially synchronized with $H(t)$ in the whole 
frequency range where $S_1$ is close to unity. The upper limit of this 
range is proportional to $\omega_2$ and therefore determined 
by $\langle \tau(H_0) \rangle$, whereas the lower limit would be 
determined by the (in our case unobservably long) $\langle \tau(0) \rangle$. 
This theme will be discussed further in Sec.~\ref{sec_a}.

\subsection{Characteristic time of the RTDs}
\label{subsec_chartime}

For each of the RTDs shown in Fig.~\ref{rtdSD},
the size of the peaks decreases for large residence times.
The rate of this decrease can be quantified by measuring
how the peak strengths decrease with increasing peak number.
We define the characteristic time, $\eta(\omega)$, for the RTDs
(in units of the field period, $2 \pi / \omega$) as
 \begin{equation}
 \label{eq_eta_def}
   \eta(\omega) = \frac{1}{ \ln S_{j}(\omega) - \ln S_{j+1}(\omega) }.
 \end{equation}
The value of $\eta(\omega)$ for any frequency of the field
is measured from the MC data by plotting
$\ln S_{j}$ vs.\ $j$.
The slope of a best-fit line through this data gives $-1/ \eta(\omega)$.
The inverse characteristic times calculated from the MC data
by a weighted least-squares fit are shown in Fig.~\ref{etaSD}.
The statistical uncertainty in the estimates for
the characteristic time
becomes large for the lowest and highest frequencies.
For the very lowest frequency shown, $1/R$=$0.05$, few switching events
contribute to
any of the peaks in the RTD, other than
the first peak.
Hence there are poor statistics in the $\ln S_{j}$ data for $j>1$,
resulting in a large uncertainty in the fitted slope.
For high frequencies the RTDs
contain many peaks.
Since the total number of switching events in each time series is
approximately equal, the statistics for each peak is poor.
In addition to this purely statistical error,
there is also a systematic effect due to the cutoff used to measure
the residence times.
Namely, the RTDs display {\it two} peaks for
the interval $0 < \omega \Delta / 2 \pi < 1$.
This extra weight in $S_{1}(\omega)$ introduces additional
systematic error which tends to raise the estimates of $|1/\eta(\omega)|$
for higher frequencies.

The theoretical calculation of the characteristic
time starts
by substituting Eq.~(\ref{eq_Somega}) for $S_{j}(\omega)$
into Eq.~(\ref{eq_eta_def}),
which gives
 \begin{eqnarray}
  \label{eq_eta_theory}
  \eta(\omega) 
               & = & \frac{-1}{\ln P_{\rm not}(\omega)} \ .
             \label{eq_etaderivation}
 \end{eqnarray}
The calculation of $\eta(\omega)$ is now trivial because
$P_{\rm not}(\omega)$ has already been
evaluated numerically
in Sec.~\ref{subsec_residencetime_strength}.
The solid curve in Fig.~\ref{etaSD} shows the theoretical
results from the full numerical calculation of $P_{\rm not}(\omega)$.
The theoretical result diverges at a value of $1/R \approx 1.19$,
the maximum
value of the frequency for which single-droplet switching is possible.
At this value of $1/R$, $P_{\rm not}$=$1$, and
the characteristic time diverges.
For frequency values $1/R > 1.19$ mechanisms
other than single-droplet decay might be possible, 
such as several droplets nucleating simultaneously. 
Given the length of our simulations, it is unlikely such 
behavior would be observed.
Indeed, the statistical and systematic errors discussed above
preclude accurate measurement of the characteristic time
from the MC data for $1/R \gtrsim 0.4$.

For low frequencies, both the theoretical and
the MC results for the characteristic time appear constant.
This motivates the search for an approximate,
analytic expression.
We start by casting Eq.~(\ref{eq_etaderivation})
in terms of the cumulative probability distribution,
 \begin{equation}
   \eta( \omega )
      = -\frac{1}{\ln \bar{F} ( \frac{\pi}{\omega} ) } \ .
 \end{equation}
Using the low-frequency approximation $t_{\rm g}$=$0$ gives
 \begin{equation}
    \label{eq_eta_main}	
     \eta( \omega )
       =  \frac{1}{2 \int_{0}^{ \frac{\pi}{2 \omega} } \rho (t') dt'} \ ,
 \end{equation}
where the upper limit in the integral is rewritten
because the nucleation rate 
$\rho(t)$ is symmetric about $t$=$\pi / 2 \omega$.
Substituting $u$=$\sin (\omega t)$ into Eq.~(\ref{eq_rho(t)}) for $\rho (t)$
allows one to write the integral above as
 \begin{equation}
  \nonumber
   \eta( \omega)
       =   \Biggl [2 \rho_{0}
       \frac{1}{\omega}
       \int_{0}^{1}
       \frac{u^3}{\sqrt{1-u^2}}
       \exp
       \biggl [ - \frac{\Xi_{0}(T)}{H_{0}} 
       \Bigl ( \frac{1}{u} -1 \Bigr ) \biggr ] du \Biggr ]^{-1} \ .
 \end{equation}
The final result for the characteristic time is given in units of
$\left < \tau \right >$ by multiplying both sides of the equation above
by $R$ and simplifying,
 \begin{equation}
 \label{eq_chartime_analytic1}
  R \eta =
    \Biggl [ \frac{\rho_{0} \left < \tau \right > }{\pi}
    \int_{0}^{1}
       \frac{u^3}{\sqrt{1-u^2}}
       \exp
       \biggl [ - \frac{\Xi_{0}(T)}{H_{0}} 
       \Bigl ( \frac{1}{u} -1 \Bigr ) \biggr ] du \Biggr ]^{-1} \ ,
 \end{equation}        
where all the quantities are known except for
the integral, which can be calculated numerically.
This theoretical result is shown in Fig.~\ref{etaSD}
as a horizontal dashed line.
The definition of $\eta (\omega)$ in Eq.~(\ref{eq_eta_def})
admits to an interpretation of the characteristic time as
an exponential decay constant for $S_{j}(\omega)$ as a function of $j$.
Therefore, $\eta (\omega)$ is a convenient average measure of an
RTD, with long characteristic times corresponding to large peaks
at long residence times. 
The exponential nature of the decay of the RTD with time should 
give rise to a Lorentzian component in the power spectral density 
with half width 
$f_{1/2} 
\approx \left[ 2 \pi R \eta \langle \tau \rangle \right]^{-1}
\approx 1.4 \times 10^{-5}$MCSS$^{-1}$,
obtained from a numerical
evaluation of Eq.~(\ref{eq_chartime_analytic1}).
This low-frequency component should be visible in the spectrum 
when $f_{1/2} < \omega/2 \pi$. 

The result in Eq.~(\ref{eq_chartime_analytic1})
was obtained in a
low-frequency approximation by setting $t_{\rm g}$=$0$.
Further approximation can be made in the evaluation
of Eq.~(\ref{eq_chartime_analytic1}).
The main contribution to the integral in the denominator occurs
for $u \approx 1$.
Expanding each factor in the integrand
in $\epsilon$=$1-u$ for small $\epsilon$ and
substituting $x^{2}$=$\Xi_{0}(T)\epsilon /H_{0}$ gives
the Gaussian approximation,
 \begin{eqnarray}
 \nonumber
  R \eta
   & = & \left [ \frac{\rho_{0} \left < \tau \right >}{\pi}
            \sqrt{\frac{2 H_{0}}{ \Xi_{0}(T)}}
            \int_{0}^{\sqrt{ \frac{\Xi_{0}(T)}{H_{0}} }} 
            e^{-x^{2}} dx \right ]^{-1} \\
   & = & \left [ \frac{\rho_{0} \left < \tau \right >}{\pi}
            \sqrt{\frac{\pi H_{0}}{2 \Xi_{0}(T)}}
            {\rm Erf}\sqrt{\frac{\Xi_{0}(T)}{H_{0}}} \right ]^{-1} \ ,
 \end{eqnarray}
which gives
$\left[ 2 \pi R \eta \left < \tau \right > \right ]^{-1}$
=$1.9 \times 10^{-5} {\rm MCSS}^{-1}$,
and was obtained by inserting quantities
from Table \ref{table_constants}.
The agreement between this value and the data
is not as good as that obtained directly from the numerical evaluation of
Eq.~(\ref{eq_chartime_analytic1}).
However, the expansion in small $\epsilon$
improves for large $\Xi_{0}(T)/H_{0}$.
So this analytic formula for $\eta (\omega)$ would
be appropriate for low frequencies and small amplitudes
of the external field, i.e. field amplitudes which
place the system even deeper into the SD region.

\section{Power spectral densities}
\label{sec_psd}
A standard method used to characterize a time
series is to calculate its power spectral density (PSD).
Figure \ref{psdSD} shows the PSDs of the raw data,
short segments of which are shown
in Fig.~\ref{mtSD}.
These PSDs are calculated with a standard FFT
algorithm implemented using a Welch window \cite{pres92}.
To reduce the variance in the PSDs,
each time series is split into several segments.
The data is then overlapped in such a way as to
obtain an ``average'' over all the
segments for each frequency bin.
The details of this method,
which we refer to as ``smoothing,''
are given in Ref.~\cite{pres92}.
Depending on the number of segments into which the original
time series is split, there is a trade-off
between frequency resolution and variance reduction.
If the time series is split into many segments,
the variance per bin is decreased at the expense of lower
frequency resolution.
Conversely, by partitioning the data into fewer long
segments one obtains higher frequency resolution and a
smaller low-frequency cutoff, at the expense of a larger variance
per bin.

The PSDs for different driving frequencies are shown in
Fig.~\ref{psdSD}.
The spectra in the main part of the plot have been
shifted in the vertical direction by arbitrary offsets.
The spectra in the inset of Fig.~\ref{psdSD}
are plotted with no offset.
Different amounts of smoothing
have been used for the low- and high-frequency regimes.
In the low-frequency regime, less smoothing is used.
This increases the frequency resolution, enabling one
to see sharper peaks in the spectra and sample
more of the low-frequency response.
In the high-frequency regime, more smoothing is used.
Since there is less power in the PSDs for the higher
frequencies, a reduction in the variance is a higher
priority than frequency resolution.
The fourth spectrum shown in Fig.~\ref{psdSD},
labeled ``background,''
corresponds to thermal fluctuations in a single-phase system.
To obtain this spectrum, a simulation was performed
on a system with the same size, temperature, and for
the same number of MCSS as the other spectra, in
a {\it static} field of $H_{0}/\sqrt{2}$.
The magnetization quickly relaxes to equilibrium.
The equilibrium fluctuations are purely thermal, and
their time correlations are exponential with a short
correlation time of only a few MCSS.
The PSD should then 
have the functional form of a Lorentzian.
When plotted on a log-log scale a Lorentzian
appears flat at low frequencies, then crosses
over into a linear curve with a slope of $-2$.
The tail on the observed noise background does not appear to
have slope=$-2$.
However this is most likely due to aliasing effects
near the Nyquist frequency, $\Omega_{N}$=$\pi$ \cite{pres92},
which is only about one order of magnitude larger than the inverse
correlation time.

To describe the PSD for each frequency, we identify four
distinct regions:
1.) the thermal noise region,
2.) the peaks,
3.) the low-frequency region,
and
4.) the intermediate region.
The inset in Fig.~\ref{psdSD} shows how the PSDs collapse onto the
background spectrum in the thermal noise region
for large frequencies.
The timescale of the thermal fluctuations is much shorter than the
shortest period of the external field.

The most prominent features of the PSDs are the sharp peaks.
For each driving frequency, the first peak in the spectrum is located
at $\omega$, the frequency of the external field.
The second peak is located at $3 \omega$, the third
peak is located at $5 \omega$, and so on.
These odd harmonic peaks arise because the shape of the time
series strongly resembles a square wave (refer to Fig.~\ref{mtSD}).
The power $p_{n}$ contained in the $n$th component of the Fourier series
for a pure square wave is
$p_{n} = 16 [\sin (n \pi/2)]^4/(n \pi)^{2}$,
which is non-zero only for odd $n$ and decays as $n^{-2}$.
The zeros in $p_{n}$ for even $n$ survive as zeros in the
PSD for a switching process in which the switching probability
density $P(t)$ reduces to a delta function \cite{shne94-1}.
The finite width of $P(t)$ for the present system smoothes out such
singularities in the PSD.
However, a clear dip in the PSD at twice the driving frequency
can be seen for $R$=$10$.
This effect was also noticed by Zhou and Moss in the weak-noise
regime for analog simulations of a bistable circuit \cite{zhou90}.

The low-frequency region comprises the portion of
each spectrum between the first peak and the lowest resolved frequency.
The PSD in this region exhibits a strong dependence on the
frequency of the external field.
For $R$=$100$, the average intensity in the low-frequency
region is approximately constant and is
weaker than the fundamental peak intensity by
about three orders of magnitude.
For $R$=$2.5$, the slope of the PSD in the low-frequency region
is close to $-2$ over almost two orders of magnitude and
contains components comparable in intensity to that of the fundamental peak.
This significant amount of power at low frequencies
is a consequence of the long residence times, i.e.\ those
residence times longer than the period of the field. 
The Lorentzian 
half width predicted from the RTDs in Sec.~\ref{subsec_chartime}, 
$f_{1/2} \approx 1.4 \times 10^{-5}$MCSS$^{-1}$, is in good agreement 
with the PSD for $R$=2.5. One can see from Fig.~\ref{psdSD} that
if the first peak for a PSD is located at a frequency
smaller (larger)
than approximately $1.4 \times 10^{-5} {\rm MCSS}^{-1}$, there
are small (large) low-frequency components.
In an analogous fashion, the low-frequency behavior of the PSDs
can be deduced by comparing the inverse characteristic time,
$\left[ 2 \pi R \eta \left < \tau \right > \right]^{-1}$,
in Fig.~\ref{etaSD} with the frequency, $\left < \tau \right >/R$
(in units of ${\rm MCSS}^{-1}$),
which is shown as a short-dashed line in the same figure.
For low driving frequencies,
$\left < \tau \right >/R < \left[ 2 \pi R \eta
\left < \tau \right > \right]^{-1}$.
Therefore, most of
the residence times are less than a period of the external field,
which corresponds to small low-frequency components in the PSDs.
For high driving frequencies,
$\left < \tau \right >/R > \left[ 2 \pi R \eta
\left < \tau \right > \right]^{-1}$,
and large low-frequency components appear in the PSDs.


The intermediate region is the portion of each spectrum
between the highest-order visible
peak and the thermal-noise region.
This region is discussed last because it is best understood
as a crossover from the peaks
to the thermal-noise region.
The structure of the odd-harmonic peaks is easiest
to discern for $R$=$100$, where the first five peaks are clearly visible.
The higher-frequency harmonics cannot be seen for two reasons:
the peak positions
become very closely spaced
on a logarithmic scale,
and the intensity of the peaks becomes
too small to be seen above the fluctuations in the PSD.
So the intermediate region
may be thought of as an envelope of the high-frequency,
odd-harmonic peaks.
A log-log plot of $p_{n}$ vs.\ odd $n$ yields a line with a slope of $-2$.
However, the portion of the PSD just to the left
of the thermal-noise region is decreasing with
a slope steeper than $-2$.
This sharp falloff is a consequence of the finite growth
time of a critical droplet.
A finite growth time effectively ``smoothes out'' the
sharp corners of the square-wave response of the magnetization,
which introduces a cutoff for the highest-frequency
components of $p_{n}$ in the intermediate region.

\section{Hysteresis loops, correlation, and period-averaged magnetization}
\label{sec_qa}

To characterize the behavior of
an entire time series we calculate the following
integrals of the magnetization,
  \begin{eqnarray}
\label{eq_A}
   A & = & -\oint m(H) \ dH \\
\label{eq_B}
   B & = & \frac{\omega}{2 \pi} \oint m(t) \ H(t) \ dt \\
\label{eq_Q}
   Q & = & \frac{\omega}{2 \pi} \oint m(t) \ dt \ ,
  \end{eqnarray}
\noindent
where
$A$ is the hysteresis-loop area,
$B$ is the correlation between the external field and
the system magnetization, and
$Q$ is the period-averaged magnetization.
These quantities are calculated over
each period in the entire time series.
From the resulting ``filtered'' time series we
construct histograms to obtain
the probability densities of $A$, $B$, and $Q$
for each separate frequency of the external field.

In Sec.~\ref{sec_a} we compare our MC data and theoretical
calculations of $A$ and $B$.
In particular, we comment on the low-frequency power-law
scaling for $A$ which has been put forth in numerous studies.
At the end of this section we show our results for $B$
and identify $A$ and $B$ as components of a nonlinear
response function.
In Sec.~\ref{sec_q} we discuss $Q$ and the absence of a
dynamic phase transition in the SD region for this system.

 \subsection{Hysteresis loops and correlation}
 \label{sec_a}
The hysteresis-loop area, $A$, represents the energy dissipated
during a single period of the applied field.
It is therefore one of the most important
physical quantities characterizing hysteretic systems,
and it is frequently measured in experiments.
Under the conditions of stochastic magnetization reversal,
$A$ and $B$ are random variables with nontrivial statistical properties.
Figure \ref{aSD} shows the probability densities of $A$.
For all values of $R \lesssim 10$, there is a sharp peak near zero.
This peak is denoted as $A_{0}$ and
corresponds to the field cycles
during which $m(t)$ does not
switch sign, but merely fluctuates near $\pm m_{\rm eq}$.
The second peak, $A_{1}$, is located near $A/(4H_{0})$=$0.5$.
It represents field cycles during which
$m(t)$ switches sign {\it once}.
The third peak, $A_{2}$, located near
a loop area of $A/(4H_{0})$=$1.0$,
represents cases when
$m(t)$ switches sign twice within the same period.
When the period of the field increases,
the weight in the peaks moves from low to high
values of $A$.
For $H(t)$ with longer periods,
the magnetization has a higher probability of
switching once or twice during a single period,
thus transferring more weight 
to the peaks $A_{1}$ and $A_{2}$.
For very low frequencies, the magnetization almost always switches
twice in each period, giving loops of type $A_{2}$.
At the same time, switching occurs earlier in each half-period.
This reduces the average loop area, giving rise to the type
of pdf shown for $R$=$100$ in Fig.~\ref{aSD}.
Figure \ref{aSD_loops} shows typical hysteresis
loops with values of $A$
corresponding to the three peaks, $A_{0}$, $A_{1}$, and $A_{2}$.

The fraction of events contained in each of these peaks
we denote as
$f_{0}$, $f_{1}$, and $f_{2}$.
These are the probabilities that the magnetization
switches zero, one, or two times in a full period of the field.
They depend on the probability that the sign of $m(t)$ is
opposite that of $H(t)$ immediately after $H(t)$ has switched sign
at the beginning of the period.
We call this phase probability $p_{-}$.
In terms of $p_{-}$ and $P_{\rm not}(\omega)$ we have:
 \begin{mathletters}
 \label{eq_fractions1}
  \begin{eqnarray}
   f_{0} & = & P_{\rm not}(\omega) \\
   f_{1} & = & p_{-} P_{\rm not}(\omega)(1-P_{\rm not}(\omega)) +
               (1-p_{-})(1-P_{\rm not}(\omega)) \\
   f_{2} & = & p_{-}(1-P_{\rm not}(\omega))^{2} \ .
\end{eqnarray}
\end{mathletters}
The phase whose probability is given by $p_{-}$ changes in a two-state
Markov process described by a transition matrix fully determined
by $P_{\rm not}(\omega)$.
It is easy to show that the stationary value of
$p_{-}$=$1/(1+P_{\rm not}(\omega))$ \cite{side98_thesis}.
After the phase distribution has reached its stationary
state, we therefore get
 \begin{mathletters}
 \label{eq_fractions2}
  \begin{eqnarray}
   f_{0} & = & P_{\rm not}(\omega) \\
   f_{1} & = & \frac{2 P_{\rm not}(\omega)(1-P_{\rm not}(\omega))}
                    {(1+P_{\rm not}(\omega))} \\
   f_{2} & = & \frac{(1-P_{\rm not}(\omega))^{2}}
                    {(1+P_{\rm not}(\omega))} \ .
\end{eqnarray}
\end{mathletters}
These expressions satisfy the
normalization condition $f_{0} + f_{1} + f_{2}$=$1$.
The $f_{i}$ are directly related to the RTD peak strengths,
$S_{j}$, through their common dependence on $P_{\rm not}$.
In particular,
$f_{0}$=$P_{\rm not}$=$1-S_{1}$.

It is less
clear how to separate the peaks when measuring the fractions from the
MC data.
For example, with cutoffs at $0.333$ and $0.666$
an event is considered
part of peak $A_{0}$ if $0 < A/4H_{0} < 0.333$,
part of peak $A_{1}$ if $0.333 < A/4H_{0} < 0.666$,
and part of peak $A_{2}$ if $0.666 < A/4H_{0} < 1.0$.
Different cutoffs were tried and judged
by how well they fit the theoretical results.
No particular choice produced a fit that
was qualitatively better than others.
Comparisons of the $f_{i}$ calculated from theory and simulation
are shown in Fig.~\ref{aSD_fracs}.
The data points are measured from the loop-area distributions,
using cutoffs at $0.333$ and $0.666$.
The solid curves are calculated from Eqs.~(\ref{eq_fractions2}a-c),
using the theoretical values for $P_{\rm not}(\omega)$.
Considering that no free parameters are used in the theory,
it agrees well with the simulation data.

The average loop area for a specific frequency is a quantity often
displayed in experimental and numerical studies of hysteresis.
It can also be obtained in the present case, and we do so below.
However, for stochastic hysteresis
this is not a very useful quantity at higher frequencies, where 
the distributions are trimodal. 
The means of the distributions in Fig.~\ref{aSD} are shown
in Fig.~\ref{aSD_stats}.
Note that the vertical bars are {\it not} error bars, but rather
give the standard deviations of the distributions.
The dotted curve is obtained by assuming that the fluctuations
of the loop areas are small in each of the three peaks.
An approximate value for the scaled loop area can be assigned to each
of the three peaks,
$\left < A_{0} \right>/4H_{0} \approx 0.00$,
$\left < A_{1} \right>/4H_{0} \approx m_{\rm eq}/2$,
and
$\left < A_{2} \right>/4H_{0} \approx m_{\rm eq}$.
With these choices we assume that the magnetization
switches when the absolute value of the external field is close to
a maximum.
The average loop area is calculated by weighting these values
by the fraction of events in each peak,
 \begin{equation}
 \label{eq_areaSDhigh}
 \frac{\left < A \right >_{\rm HF}}{4H_{0}} =
     f_{0}(\omega) \Biggl (\frac{\left < A_{0} \right >}{4H_{0}} \Biggr ) +
     f_{1}(\omega) \Biggl (\frac{\left < A_{1} \right >}{4H_{0}} \Biggr ) +
     f_{2}(\omega) \Biggl (\frac{\left < A_{2} \right >}{4H_{0}} \Biggr  )
 \end{equation}
where ``HF'' stands for ``high-frequency.''
This expression breaks down in the low-frequency limit, as
$f_{2} \rightarrow 1$ and 
$\left < A_{2} \right>$ becomes strongly frequency dependent.
Next we discuss this effect.

For any finite time series
there is a sufficiently low frequency such that
the magnetization always switches during every half-period of
the field.
For very low frequencies the magnetization switches, on average,
{\it before\/} the field reaches its extreme value during every half-period.
Therefore, the loop-area distribution becomes unimodal,
and the mean loop area decreases with decreasing frequency.
Since the probability density of switching times, $P(t)$,
is narrow and unimodal for low frequencies
(refer to Fig.~\ref{fig_rho}),
we can choose the typical time when the magnetization switches during
each half period, $t_{\rm s}$, to be either its mean, mode, or median.
We use the latter because it
is analytically more tractable and has an
analytic asymptotic approximation.
To determine the median of $P(t)$ we must solve $F(t_{\rm s})$=$1/2$,
where the cumulative probability distribution $F(t)$ is 
given by Eq.~(\ref{eq_F(t)growth2}).
To simplify the form of this equation we note that for
low frequencies the hysteresis loops are square
(refer to Fig.~\ref{aSD_loops}(d)),
which means that the growth time becomes small compared to $t_{\rm s}$
and can be ignored.
Thus, the equation for $t_{\rm s}$ becomes
 \begin{equation}
  F(t_{\rm s}) = 1 - \exp \left [-\int^{t_{\rm s}}_{0} \rho(t') dt' \right ] 
= \frac{1}{2} \ .
 \end{equation}
After rearranging and substituting
$x$=$\Xi_{0}/[H_{0} \sin (\omega t)]^{d-1}$,
we obtain
 \begin{eqnarray}
 \ln 2 & = & \int^{t_{\rm s}}_{0} \rho(t') dt' \\
       & = &  \rho_{0}
              \frac{ e^{\frac{\Xi_{0}(T)}{H_{0}^{d-1}}} }{H_{0}^{K}}
              \int^{t_{\rm s}}_{0}
              [H_{0} \sin(\omega t')]^{K}
              e^{-\frac{\Xi_{0}(T)}{[H_{0} \sin (\omega t')]^{d-1}}} dt' \\
       \label{eq_hsFull}
       & = & \rho_{0}
      \frac{ e^{\frac{\Xi_{0}(T)}{H_{0}^{d-1}}} \Xi_{0}^{\frac{K+1}{d-1}} }
                   { H_{0}^{K+1} (d-1) \omega }
              \int^{\infty}_{\frac{\Xi_{0}}{H_{\rm s}^{d-1}}}
              \frac{x^{-\frac{K+d}{d-1}} e^{-x}}
      {\sqrt{1-\Bigl [ \frac{1}{H_{0}} (\frac{\Xi_{0}}{x})^\frac{1}{d-1} 
                            \Bigr ]^{2}}}   dx \ .
 \end{eqnarray}
As the frequency is decreased the switching field,
$H_{\rm s}$=$H_{0} \sin (\omega t_{\rm s})$, decreases.
For very low frequencies $x \gg \Xi_{0}/H_{0}^{d-1}$
and the radical in Eq.~(\ref{eq_hsFull}) is of order one,
resulting in the following expression
 \begin{equation}
  \nonumber
  \ln 2 = \rho_{0}
   \frac{ e^{\frac{\Xi_{0}(T)}{H_{0}^{d-1}}} \Xi_{0}^{\frac{K+1}{d-1}} }
                   { H_{0}^{K+1} (d-1) \omega }
              \int^{\infty}_{\frac{\Xi_{0}}{H_{\rm s}^{d-1}}}
              x^{-\frac{K+d}{d-1}} e^{-x} dx \ ,
 \end{equation}
which may also be derived through a
linear approximation to the sinusoidal field,
$H(t) \approx H_{0} \omega t$.
After rearranging and simplifying this becomes \cite{side98-MMM} 
 \begin{equation}
  \label{eq_gammaloop}
    C H_{0} \omega
      = \Gamma \left ( 1-\frac{K+d}{d-1},\frac{\Xi_{0}}{H_{\rm s}^{d-1}}
         \right ) \,
 \end{equation}
where $\Gamma(a,x)$ is the incomplete gamma function \cite{abra70},
and we define
 \begin{equation}
  \label{eq_constant_forloop}
   C =
     \frac{ \ln 2 \ H_{0}^{K} (d-1) e^{-\frac{\Xi_{0}(T)}{H_{0}^{d-1}}}
            \Xi_{0}^{-\frac{K+1}{d-1}}}
          { \rho_{0} } .
 \end{equation}
With $d$=$2$, $K$=$3$, and
the values found in Table \ref{table_constants},
$C$=$0.101 \ J^{-1} \ {\rm MCSS}$.
For small $\omega$ the hysteresis loops are practically square,
so the scaled loop area in the low-frequency (LF) limit
can be expressed as
\begin{equation}
\label{eq_loopLF}
\nonumber
 \frac{\left < A \right >_{\rm LF}}{4H_{0}} 
\approx m_{\rm eq}\frac{H_{\rm s}(\omega)}{H_{0}}.
\end{equation}
The switching field $H_{\rm s}(\omega)$ 
is obtained from a numerical solution of
Eq.~(\ref{eq_hsFull}), and the result
for $\left < A \right >_{\rm LF}/4H_{0}$ is shown as the
solid curve in Fig.~\ref{aSD_stats}(a)-(b).
Figure \ref{aSD_stats}(b) shows the good agreement between
this parameter-free calculation and the MC data
for frequencies $1/R \leq 0.05$.

One can obtain an approximate analytic solution by taking the
first term in the
asymptotic expansion \cite{abra70}
\begin{equation}
 \Gamma(a,x) \sim x^{a-1} e^{-x} \Bigl [ 1 + \frac{a-1}{x} + \ldots \Bigr ]
\end{equation}
for large $x$.
By ignoring the factor $x^{a-1}$, one obtains
a completely analytic solution for $H_{\rm s}$ in the extreme LF limit,
resulting in the asymptotic,
logarithmic frequency dependence for the loop area
 \begin{equation}
 \label{eq_SDloopanalytic}
 \left < A \right >_{\rm LF} \approx
    4 \Xi_{0}^{\frac{1}{d-1}}
    \Bigl [- \ln (C H_{0} \omega) \Bigr ]^{-\frac{1}{d-1}} \ .
 \end{equation}

In Fig.~\ref{aSD_stats}
we show calculations of both this asymptotic analytic
result (short-dashed curve) for the loop area
and the loop area obtained from the numerical solution (solid curve) of
Eq.~(\ref{eq_hsFull}).
The dashed lines in Fig.~\ref{aSD_stats}(b) are linear least-squares fits
to the full numerical solution over almost four decades in frequency.
The slopes of these fit lines would give an ``effective'' exponent b for 
the loop area,
$A \propto \omega^{b}$.
For the frequency regime shown,
these effective power-law exponents for the loop area depend on
frequency.
These results show no evidence of an
overall power-law relationship between the frequency
and the loop area.

To the best of our knowledge, the present study is the first
to give a complete solution of $A$ and emphasize its
possible significance for the low-frequency power-law
behavior of $A$ reported in the literature.
This issue is discussed in further detail
in a separate paper \cite{side98-MMM}.
Theoretical considerations of nucleation effects
on hysteresis have been mentioned previously in the literature
\cite{beal94,rao90,thom93},
reporting the same asymptotic frequency behavior
for the loop area as in Eq.~(\ref{eq_SDloopanalytic}).
However, we stress that the purely logarithmic dependence of the loop area
on frequency and amplitude of the external field will approach
the exact solution obtained from Eq.~(\ref{eq_gammaloop}) only
for {\it extremely} low frequencies,
as illustrated by the poor agreement between the short-dashed
and solid curves in Fig.~\ref{aSD_stats}(b).
The units of frequency used in our simulations may be converted roughly
into Hertz by  equating a phonon frequency
with an inverse Monte Carlo step per spin, $\bar{\nu}$=$1 \ {\rm MCSS}^{-1}$,
with $\bar{\nu}$=$10^{9}$--$10^{12} {\rm Hz}$ as a reasonable estimate.
Then, the lowest frequency calculated for the solid curve
in Fig.~\ref{aSD_stats}(b) corresponds to a field period on the order of
thousands to millions of years.
The full asymptotic behavior of the loop area is realized for frequencies
lower still, and is therefore well outside the range of feasible experiments.
However, the lowest-frequency MC data point corresponds to a field period
on the order of microseconds to seconds, suggesting the low-frequency
SD behavior of the loop area described by our full numerical calculation
could be experimentally observable.

In extensive simulations, Acharyya and Chakrabarti
have studied the hysteretic response of Ising systems
with respect to
changes in field amplitude and frequency, temperature, system size, and
system dimension  \cite{acha94_review}.
In particular, they have presented power-law scaling relations for $A$.
However, their simulations in $d$=$2$ are all done with very large
field amplitudes.
For the temperatures used in their papers, these field
amplitudes place their system in a regime
we refer to as the ``strong-field (SF) region,''
where the size of a critical droplet becomes comparable to the
lattice constant $a$, and the droplet picture of metastable
decay breaks down \cite{rikv94}.
The investigation of hysteresis in
kinetic Ising models by Lo and Pelcovits \cite{lo90}
is not restricted to the SF region.
In particular, the range of field amplitudes used
places their simulations
in the SD, MD and SF regions as the amplitude is increased.
The hysteretic response, including the loop area,
requires qualitatively different theoretical descriptions in each
of these regions.
This drastic change in the Ising model contrasts with
a coherent rotation model of magnetization reversal,
for which increasing the
field strength would merely decrease the energy barrier that
the system must overcome
for the magnetization to be aligned with the field.
For the kinetic Ising model and other spatially extended systems
however,
increasing the field not only decreases
the free-energy barrier for forming a single
critical droplet of the stable phase,
it also changes the reversal mode.
Our results emphasize the importance of a knowledge of the decay mode
in order to obtain the correct frequency
dependence of the average loop area over a wide range of frequencies.

Mahato and collaborators \cite{maha94,maha97} also considered
a driven bistable
system in the presence of noise.
From the RTDs they calculate what they 
describe as an ``average hysteresis loop,'' and they propose 
a maximum in this quantity with respect to noise strength as a 
manifestation of SR. 
We find their definition of a loop area rather unphysical, except 
for very low frequencies. However, the role of the hysteresis-loop area 
as a measure of energy dissipation indicates that a maximum 
in $A$ may be considered an aspect 
of SR, which we suggest could be termed ``stochastic energy resonance.''

Relatively few studies have considered $B$, the correlation of the
magnetization with the external field.
For Ising models subject to
oscillating fields,
these studies \cite{neda95,neda96} have found a
temperature at which the correlation attains
a maximum value for a given frequency, amplitude, and system size.
However, as recently pointed out by Acharyya \cite{acha98},
this feature is not a sign of SR.
These simulations are mostly performed above $T_{c}$ in a
regime where the metastable state no longer exists.
Therefore, comparison of our simulations or analytic results with
these studies is not relevant.

Our theoretical derivations of $B$ are similar to those for $A$ 
and are also given as high- and low-frequency approximations.
Figure \ref{bSD} shows our MC data for $B$ along with the theoretical
result for the high-frequency regime (dotted curve) and the
low-frequency regime (solid curve).
We assume that $H_{\rm s} \approx H_{0}$ for high frequencies,
and that the magnetization switching is stochastic with
either zero, one, or two switches every period.
The only non-zero contribution to the correlation comes
from those periods when the magnetization switches only once.
We use notation analogous to the high-frequency loop-area
calculation for each of these three contributions,
$\left < B_{0} \right>/(2H_{0}/\pi) \approx 0.0$,
$\left < B_{1} \right>/(2H_{0}/\pi) \approx -m_{\rm eq}/2$,
and
$\left < B_{2} \right>/(2H_{0}/\pi) \approx 0.0$,
with
$\left < B \right>_{\rm HF}/(2H_{0}/\pi)$
calculated as a weighted average similar to Eq.~(\ref{eq_areaSDhigh}),
 \begin{equation}
 \label{eq_corrSDhigh}
 \frac{\left < B \right >_{\rm HF}}{2H_{0}/\pi} =
     f_{0}(\omega) \Biggl (\frac{\left < B_{0} \right >}
        {2H_{0}/\pi} \Biggr ) +
     f_{1}(\omega) \Biggl (\frac{\left < B_{1} \right >}
        {2H_{0}/\pi} \Biggr ) +
     f_{2}(\omega) \Biggl (\frac{\left < B_{2} \right >}
        {2H_{0}/\pi} \Biggr  ) \ .
 \end{equation}
Figure \ref{bSD} shows that
$\left < B \right >_{\rm HF}/(2H_{0}/\pi) \rightarrow 0$ as
$\omega \rightarrow 0$ and as $\omega \rightarrow \infty$,
which is a direct consequence of the behavior of $f_{1}(\omega)$
(see Fig.~\ref{aSD_fracs}(b)).

The low-frequency calculation for $B$ is also similar to that for $A$
and assumes that the magnetization switches abruptly 
at $|H_{\rm s}| < H_{0}$ during each half-period.
This assumption together with the definition of $B$, Eq.~(\ref{eq_B}), 
yields 
 \begin{equation}
 \label{eq_corrLF}
  \frac{\left < B \right >_{\rm LF}}{2H_{0}/\pi} =
    m_{\rm eq}
    \sqrt{ 1- \Biggl ( \frac{H_{\rm s}(\omega)}{H_{0}} \Biggr )^{2} } \ .
 \end{equation}
As for $A$, the switching field $H_{\rm s}(\omega)$ 
is obtained from a numerical solution of Eq.~(\ref{eq_hsFull}).
There is good agreement with the MC data for the two lowest frequencies.
As the frequency increases the theoretical result approaches zero. 
However, the assumption that the magnetization
switches during every half-period begins to break down around
$1/R$=$0.05$, where the MC result for $B$ passes through zero. 

Although the system studied here is both stochastic and highly nonlinear, 
the physical significance of the integrals $A$ and $B$ can be clarified by 
comparison with deterministic linear response theory. In that limit one 
easily 
finds that $A/(\pi H_0^2)$ and $2B/H_0^2$ correspond to the dissipative and 
reactive parts of the complex linear response function, respectively.
It is therefore natural to combine $A$ and $B$ into an analogous 
{\em nonlinear\/} response function, 
 \begin{equation}
 \label{eq_X}
  X(H_0,T,\omega) = \frac{1}{H_0^2} \left[ 2B + \frac{i}{\pi} A \right] \;.
 \end{equation}
The maximum in $A$ and the sign change in $B$, which occur 
close together in frequency, are characteristic 
behaviors of the dissipative and a reactive parts of a response function 
near resonance. It is reasonable to associate this behavior with SR. 
However, as we pointed out at the end of 
Sec.~\ref{subsec_residencetime_strength}, $m(t)$ remains essentially 
synchronized with $H(t)$ as the driving frequency is lowered further 
below this narrow frequency range.
The system then switches reliably during every 
half-period of the field, while the switching occurs earlier 
and earlier in the half-period as the frequency is lowered. 
In this low-frequency regime, the norm of $X$ remains close to
its maximum value of $4 m_{\rm eq} / (H_0 \pi)$, and its 
phase gives a meaningful measure of the period-averaged phase lag. 
The latter increases monotonically from zero at $\omega=0$ to $\pi/2$ 
at the frequency where $B$ crosses zero. 
We believe the system should be considered as resonant in this 
whole range of low frequencies.

 \subsection{Period-Averaged Magnetization}
 \label{sec_q}

The period-averaged magnetization $Q$ has been considered as a
``dynamic order parameter'' for systems exhibiting hysteresis
\cite{tome90,acha94_review,acha92-1,acha92-2,acha94,acha94-2,lo90}.
Those studies of the Ising
model have suggested the existence of a dynamic phase
transition between
$\langle Q \rangle \neq0$ and
$\langle Q \rangle$=$0$.
As for the hysteresis-loop areas, the statistical properties
of the period-averaged magnetization in the SD region
are not well characterized simply by its mean.
Figure \ref{qSD} shows
the probability densities of $Q$ in the SD region.
For all but the very highest values of $R$, the distributions show
two sharp peaks near $Q$=$\pm m_{\rm eq}$ due to $m(t)$ oscillating
near the spontaneous magnetization during most of the field cycles.
The contributions to the $Q$ distributions near $Q$=$0$ occur when
the magnetization switches twice in one period.
The contributions to the peak
centered around $Q$=$+0.5$ occur for those periods in which
the magnetization switches only once.
Note that there is not a corresponding peak
at $Q$=$-0.5$.
This is an effect of the way we
calculate the period-averaged magnetization, which considers the beginning
of a period to start when
$H(t)$=$0$ and $\dot{H} > 0$.
We would have obtained a peak near $Q$=$-0.5$ if we had started
with
$H(t)$=$0$ and $\dot{H} < 0$.

Even by inspecting the distributions for $Q$, no dynamic phase
transition can be seen. 
While the means of the distributions for high (low) frequencies are
nonzero (zero), this happens smoothly as weight shifts from the peaks near
$Q$=$\pm 1$ to the peak at $Q$=$0$.
As we will show in a forthcoming paper, the situation is quite different
in the MD region, where we find strong evidence for a dynamic phase
transition \cite{side98-MD}.

\section{Discussion}
\label{sec_conclusion}

The mechanism by which a metastable phase decays depends
sensitively on the system size, the temperature, and the strength
of the applied field.
For small systems and weak fields, the decay proceeds through the
nucleation and growth of a {\it single} droplet of overturned spins.
This regime has been termed the single-droplet (SD) region.
In this region the magnetization response consists
of rapid transitions between two states; one with the majority of the
spins up, and one with the majority of the spins down.
The
resulting time series is well described
in terms of a Poisson process with a time-dependent rate
obtained from the nucleation rate and growth velocity of
droplets of the stable phase.
The time dependence enters the nucleation rate by replacing the constant
field $H$ by $H(t)$=$H_{0} \sin (\omega t)$.
This central idea provides the analytic framework
for theoretical descriptions
of the quantities measured from our MC simulations.
These quantities include residence-time distributions (RTDs), 
power spectral densities (PSDs), hysteresis-loop areas, and the correlation 
between the magnetization and the oscillating field. 
The agreement between all of our theoretical calculations
and the MC data is very good, especially considering that the theory
contains no adjustable parameters.
All of the constants used are either known from droplet theory
or are measured from MC simulations of field reversal in kinetic Ising models.
To the best of our knowledge,
the present study is the first which explicitly
considers hysteresis for the Ising model in the SD regime.

The frequency dependence of the RTD peak shapes
and peak strengths are calculated by numerically evaluating analytic
expressions obtained from a time-dependent extension of classical
nucleation theory.
The good agreement between our theoretical calculations and MC data 
supports the model of magnetization switching as a Poisson process with
a variable rate, given by substituting a sinusoidal field dependence
for the static field in the nucleation rate.

The frequency dependence of the RTD peak strengths, the 
hysteresis-loop areas, and the correlation between the magnetization and the 
field all indicate the presence of stochastic resonance (SR) in 
the two-dimensional Ising model in this parameter regime. 
This observation is consistent with recent studies of SR in other 
systems of coupled bistable elements, some of which pointed out the 
importance of nucleation of kink-antikink pairs to what has been 
termed array enhanced stochastic resonance (AESR). 
For any nucleation process, there should be crossovers between 
coexistence (CE), single-droplet (SD), and multidroplet (MD) types of 
behavior as a consequence of the interplay between the sizes and separations 
of the critical fluctuation(s) and the size of the system.
We believe these crossovers should be relevant to the dependence of the
amount of enhancement on the number of elements observed in
other systems exhibiting AESR as well.

We also calculate the hysteresis-loop area, $A$,
in the low- and high-frequency regimes. 
Because of its role as a measure of the energy dissipation in the system, 
this is a quantity of particular experimental significance. 
For high frequencies the loop-area distributions are
trimodal due to the stochastic switching behavior.
In this regime we calculate $\langle A \rangle $ as a weighted
average of the loop areas obtained when the magnetization
switches zero, one, or two time(s) during a period of the field.
For the low-frequency regime we obtain an
analytic expression for $\langle A \rangle $. 
Our theoretical calculation agrees well with our MC results
and predicts an {\it extremely} slow crossover to a logarithmic 
dependence of the loop area on $H_{0} \omega$.
The switching dynamics is 
dominated by nucleation and indicates no overall power-law
dependence for the loop area on field amplitude and/or frequency,
in contrast to what has been claimed in 
other simulational and experimental studies. 
However, we emphasize that numerical analysis of data generated by 
our analytic solution, even over several frequency decades, could easily 
lead to the conclusion that the data were taken from a power law. 

The period-averaged magnetization,
$Q$, has been proposed as an ``order parameter'' associated
with a dynamic phase transition in kinetic Ising models.
However, in the parameter range studied here, 
the probability densities of $Q$ show no sign of a sharp
transition as the frequency of the external field is varied.
Indeed, due to the multi-peaked nature of the distributions for intermediate
frequencies, the mean value of $Q$ is not a useful quantity in the SD region.
In the MD region the behavior of $Q$ is radically different, as we will
discuss in forthcoming papers \cite{side98-MD}.

We also computed the power spectral densities (PSDs)
from the simulated magnetization time series.
We qualitatively explain the various features of the spectra
in the full frequency range from the lowest observable frequencies 
to the rapid fluctuations due to thermal noise. 
Specifically, we make a connection between the RTDs and the 
low-frequency behavior in the PSDs through
the characteristic time of the RTDs.
Inverse characteristic times that are smaller (larger) than
the frequency of the applied field correspond to large (small)
low-frequency components in the PSDs.
Our theoretical derivation of the characteristic time also agrees well
with the characteristic times obtained from
the simulated RTDs, except at very high frequencies.
The relatively poor agreement between the MC data and the theory
for high frequencies
of the external forcing field is common to most
of the quantities measured
and is most likely due to the poor quality of the 
MC data for high frequencies.
However, it could also be a sign of
breakdown in the adiabatic approximation underlying the
assumption that the functional form of the nucleation rate
and the calculation of the growth-time corrections
do not change for high frequencies \cite{shne94-2}.

In summary,
we have studied stochastic hysteresis in the kinetic Ising model,
a spatially extended, bistable system with thermal fluctuations.  
We emphasize not only the detailed differences between
hysteresis in mean-field models and Ising models, but
also the qualitatively different
response that the Ising model displays for particular regimes
of system size and field amplitude.
Our theoretical and numerical study is the first to consider the effects of
these different decay regimes on hysteresis, which may be
relevant to the interpretation of simulational and experimental results.
Especially for certain technological applications,
an Ising system should be a good candidate to model the
behavior of ferromagnetic and ferroelectric materials in oscillating 
external fields.
Finally we note that the quantities that we have analyzed numerically could 
all be measured in experiments on hysteresis in a variety of
systems and analyzed by methods essentially identical to our analysis of 
the MC data. 

\acknowledgments{
We would like to thank
P.D Beale,
G. Brown,
M. Grant,
W. Klein,
M. Kolesik,
R.A. Ramos,
and H. Tomita
for useful discussions.
S.W.S and P.A.R appreciate hospitality and support from the
Colorado Center for Chaos and Complexity
during the 1997 Workshop on Nucleation Theory
and Phase Transitions.
Supported in part by
the FSU Center for Materials Research and Technology (MARTECH),
by the
FSU Supercomputer Computations Research Institute (SCRI) under DOE
Contract No.\ DE-FC05-85ER25000,
and by NSF
Grants No.\ DMR-9315969, DMR-9634873, and DMR-9520325.}

\appendix
\section{Derivation of the residence-time distributions}
\label{appendix_residencetime_dis}

The $\alpha^{\rm th}$ residence time
is defined as
 \begin{equation}
  \Delta^{\alpha} = t_{\uparrow}^{\alpha} + \theta^{\alpha},
 \end{equation}
where the times $t_{\uparrow}^{\alpha}$ and $\theta^{\alpha}$
are shown schematically
in Fig.~\ref{field_sch}.
We define
$t_{\uparrow}$ as the time when a switching event takes
place,
as measured from the first time at which $H(t)$=$0$,
{\it after} the previous switching event.
Without loss of generality, this time
can be set to $t$=$0$.
$\theta$ is the time from a switching event to the next
change in the sign of the external field.
This decomposition of the residence time facilitates the
calculation of the probability density functions (pdf) for both
$t_{\uparrow}^{\alpha}$ and $\theta^{\alpha}$.
When $t_{\uparrow}$ falls during the $j$=$1$ period
(see Fig~\ref{field_sch} for an explanation of the indexing scheme),
its probability density is given by Eq.~(\ref{eq_P(t)}), i.e.
$p_{\uparrow}(t_{\uparrow})$=$P(t_{\uparrow})$.
We easily generalize to the case when
$t_{\uparrow}$ falls during the $j$th period.
This is obtained by finding the probability that, given the magnetization
has {\it not} switched in the previous $j-1$ periods,
it switches at a time $t_{\uparrow}$
during the $j$th period,
 \begin{equation}
 \label{eq_Pup1}
 p_{\uparrow}(t_{\uparrow}) = [P_{\rm not}(\omega)]^{j-1}
                              P \left [ t_{\uparrow}
                              -(j-1) \frac{2 \pi}{\omega} \right ] \ .
 \end{equation}
The pdf for $\theta$,
$p_{\theta}(\theta)$, is calculated from
$p_{\uparrow}(t_{\uparrow})$ by using the fact that
$\theta^{\alpha}$ and $\theta^{\alpha + 1}$ should
be independent and identically distributed.
Then the following substitution holds,
 \begin{equation}
 \label{eq_tup_substitution}
 t_{\uparrow}^{\alpha} 
 + \theta^{\alpha+1} = (j- \frac{1}{2}) \frac{2 \pi}{\omega},
 \end{equation}
which gives
 \begin{eqnarray}
  p_{\theta}(\theta^{\alpha})
   \nonumber
    & = &  p_{\theta}(\theta^{\alpha+1}) \\
   \nonumber
    & = & \sum_{j=1}^{\infty} p_{\uparrow} \left [
          (j-\frac{1}{2}) \frac{2 \pi}{\omega} - \theta^{\alpha+1} \right ] \\
   \nonumber
    & = & \sum_{j=1}^{\infty} [P_{\rm not}(\omega)]^{j-1} P \left [ 
          (j-\frac{1}{2}) \frac{2 \pi}{\omega} - \theta^{\alpha} -
          (j-1)\frac{2 \pi}{\omega} \right ] \\
    & = & P \left ( \frac{\pi}{\omega} - \theta^{\alpha}
          \right ) \sum_{j=1}^{\infty} P_{\rm not}(\omega)^{j-1}.
 \end{eqnarray}
Thus,
 \begin{equation}
 p_{\theta}(\theta)
     =  P \left ( \frac{\pi}{\omega} - \theta 
                            \right ) 
            \Bigl [1 - P_{\rm not}(\omega) \Bigr ]^{-1}.
  \end{equation}
The pdf of the total residence time, $\Delta$=$t_{\uparrow} + \theta$,
is given by the convolution of the pdfs of each term:
 \begin{mathletters}
 \label{eq_pdelta1}
  \begin{eqnarray}
   \Pi(\Delta) 
    & = & \int_{\theta_{a}(\Delta)}^{\theta_{b}(\Delta)}
          p_{\uparrow}(\Delta - \theta)
          p_{\theta}(\theta) d\theta \\
    & = & \frac{[P_{\rm not}(\omega)]^{j-1}}
          {[1-P_{\rm not}(\omega)]}
          \int_{\theta_{a}(\Delta)}^{\theta_{b}(\Delta)}
          P \Bigl [\Delta - \theta - (j-1) \frac{2 \pi}{\omega} \Bigr ]
          P \Bigl (\frac{\pi}{\omega}-\theta \Bigr ) d\theta,
  \end{eqnarray}
 \end{mathletters}
\noindent
where $j$ = $\lceil \omega \Delta/(2 \pi) \rceil$.
The notation $\lceil x \rceil$ is defined as the smallest
integer greater than $x$.
For $j$=$1$, the integration limits are given by
 \begin{mathletters}
 \label{eq_thetalimits}
  \begin{eqnarray}
  \theta_{a}(\Delta) & = & {\rm Max} 
  \left [0,\Delta-\frac{\pi}{\omega} \right ] \\
  \theta_{b}(\Delta) & = & {\rm Max} \left[0,{\rm Min}
                 [\Delta - t_{0} , \frac{\pi}{\omega}-t_{0}] \right] \ ,
  \end{eqnarray}
 \end{mathletters}
\noindent
which ensure that
the integrand in the equation for $\Pi(\Delta)$ is positive.
To implement the calculation of the RTDs, $\Pi(\Delta)$ is numerically
integrated for $j$=$1$ at 50 equally spaced values in the
interval $0 < \Delta < 2 \pi / \omega$,
to generate the first peak in the RTD.
To obtain the second peak, the values of the distribution are
shifted to the next interval $2 \pi / \omega < \Delta < 4 \pi / \omega$,
and reduced by
a factor of $P_{\rm not}(\omega)$.
This process can be repeated to obtain all of the higher-order
peaks in the RTD.


\begin{thebibliography}{10}

\bibitem[{\rm a)}]{SIDESadd}
Present address at Florida State University
E-mail: sides@scri.fsu.edu

\bibitem[{\rm b)}]{RIKadd}
Present and permanent address at Florida State University
E-mail: rikvold@scri.fsu.edu

\bibitem[{\rm c)}]{NOVadd}
Present address at Florida State University
E-mail: novotny@scri.fsu.edu

\bibitem{stei1892}
C.~P. Steinmetz, Trans.\ Am.\ Inst.\ Electr.\ Eng.\ {\bf 9},  3  (1892).

\bibitem{maye91}
I.~D. Mayergoyz, {\em Mathematical Models of Hysteresis} (Springer, New York,
  1991).

\bibitem{ahai96}
A. Aharoni, {\em Introduction to the Theory of Ferromagnetism} (Clarendon,
  Oxford, 1996).

\bibitem{ishi71}
Y. Ishibashi and Y. Takagi, J.\ Phys.\ Soc.\ Jpn.\ {\bf 31},  506  (1971).

\bibitem{orih92}
H. Orihara and Y. Ishibashi, J.\ Phys.\ Soc.\ Jpn.\ {\bf 61},  1919  (1992).

\bibitem{beal94}
P.~D. Beale, Integrated Ferroelectrics {\bf 4},  107  (1994).

\bibitem{rao91}
M. Rao and R. Pandit, Phys.\ Rev.\ B {\bf 43},  3373  (1991).

\bibitem{bard80}
A.~J. Bard and L.~R. Faulkner, {\em Electrochemical Methods: Fundamentals and
  Applications} (Wiley, New York, 1980).

\bibitem{rikv95}
P.~A. Rikvold {\it et~al.}, Surf.\ Sci. {\bf 335},  389  (1995).

\bibitem{chen96}
A. Cheng and M. Chaffrey, J.\ Phys.\ Chem. {\bf 100},  5608  (1996).

\bibitem{nayf95}
A.~H. Nayfeh and B. Balachandran, {\em Applied Nonlinear Dynamics} (Wiley, New
  York, 1995).

\bibitem{visi94}
A. Visintin, {\em Differential Models of Hysteresis} (Springer, Berlin, 1994).

\bibitem{brok96}
M. Brokate and J. Sprekels, {\em Hysteresis and Phase Transitions} (Springer,
  New York, 1996).

\bibitem{mart87}
Y. Martin and H. Wickramasinghe, Appl.\ Phys.\ Lett.\ {\bf 50},  1455  (1987).

\bibitem{chan93}
T. Chang, J.~G. Zhu, and J. Judy, J.\ Appl.\ Phys.\ {\bf 73},  6716  (1993).

\bibitem{lede93}
M. Lederman, G. Gibson, and S. Schultz, J.\ Appl.\ Phys.\ {\bf 73},  6961
  (1993).

\bibitem{lede94}
M. Lederman, D. Fredkin, R. O'Barr, and S. Schultz, J.\ Appl.\ Phys.\ {\bf 75},
   6217  (1994).

\bibitem{lede94-2}
M. Lederman, S. Schultz, and M. Ozaki, Phys.\ Rev.\ Lett. {\bf 73},  1986
  (1994).

\bibitem{rich95}
H.~L. Richards, S.~W. Sides, M.~A. Novotny, and P.~A. Rikvold, J.\ Magn.\
  Magn.\ Mater.\ {\bf 150},  37  (1995).

\bibitem{rich96}
H.~L. Richards, M.~A. Novotny, and P.~A. Rikvold, Phys.\ Rev.\ B {\bf 54},
  4113  (1996).

\bibitem{rich96-2}
H.~L. Richards {\it et~al.}, Phys.\ Rev.\ B {\bf 55},  11521  (1997).

\bibitem{kole97}
M. Kolesik, M.~A. Novotny, and P.~A. Rikvold, Phys.\ Rev.\ B. {\bf 56},  11791
  (1997).

\bibitem{kole97-MRS}
M. Kolesik, M.~A. Novotny, and P.~A. Rikvold, Mater.\ Res.\ Soc.\ Conf.\ Proc.\
  Ser.  (1997).

\bibitem{he93}
Y. He and G. Wang, Phys.\ Rev.\ Lett.\ {\bf 70},  2336  (1993).

\bibitem{rao89}
M. Rao, H. Krishnamurthy, and R. Pandit, J.\ Phys.\ C {\bf 1},  9061  (1989).

\bibitem{rao90}
M. Rao, H. Krishnamurthy, and R. Pandit, Phys.\ Rev.\ B {\bf 42},  856  (1990).

\bibitem{rao90_2}
M. Rao, H. Krishnamurthy, and R. Pandit, J.\ Appl.\ Phys. {\bf 67},  5451
  (1990).

\bibitem{jiang95}
Q. Jiang, H.-N. Yang, and G.-C. Wang, Phys.\ Rev.\ B {\bf 52},  14911  (1995).

\bibitem{benz81}
R. Benzi, A. Sutera, and A. Vulpiani, J.\ Phys.\ A {\bf 14},  L453  (1981).

\bibitem{adi96}
A.~R. Bulsara and L. Gammaitoni, Physics Today {\bf 49},  Issue 3, 39  (1996).

\bibitem{jung90}
P. Jung, G. Gray, and R. Roy, Phys.\ Rev.\ Lett. {\bf 65},    (1990).

\bibitem{tome90}
T. Tom{\'e} and M.~J. de~Oliveira, Phys.\ Rev.\ A {\bf 41},  4251  (1990).

\bibitem{luse94}
C. Luse and A. Zangwill, Phys.\ Rev.\ E {\bf 50},  224  (1994).

\bibitem{loch96}
M. L\"{o}cher, G. Johnson, and E. Hunt, Phys.\ Rev.\ Lett. {\bf 77},  4698
  (1996).

\bibitem{benz85}
R. Benzi, A. Sutera, and A. Vulpiani, J.\ Phys.\ A {\bf 18},  2239  (1985).

\bibitem{lind95}
J.~F. Lindner {\it et~al.}, Phys.\ Rev.\ Lett. {\bf 75},  3  (1995).

\bibitem{lind96}
J.~F. Lindner {\it et~al.}, Phys.\ Rev.\ E {\bf 53},  2081  (1996).

\bibitem{marc96}
F. Marchesoni, L. Gammaitoni, and A. Bulsara, Phys.\ Rev.\ Lett. {\bf 76},
  2609  (1996).

\bibitem{brey96}
J.~J. Brey and A. Prados, Phys.\ Lett.\ A {\bf 216},  240  (1996).

\bibitem{neda95}
Z. N\'{e}da, Phys.\ Rev.\ E {\bf 51},  5315  (1995).

\bibitem{neda96}
Z. N\'{e}da, Phys.\ Lett.\ A {\bf 210},  125  (1996).

\bibitem{pras97}
P.~M. Gade, R. Rai, and H. Singh, Phys.\ Rev.\ E {\bf 56},  2518  (1997).

\bibitem{jung92}
P. Jung, U. Behn, E. Pantazelou, and F. Moss, Phys.\ Rev.\ A {\bf 46},  R1709
  (1992).

\bibitem{glau63}
R. Glauber, J.\ Math.\ Phys.\ {\bf 4},  294  (1963).

\bibitem{rikv94}
P.~A. Rikvold, H. Tomita, S. Miyashita, and S.~W. Sides, Phys.\ Rev.\ E {\bf
  49},  5080  (1994).

\bibitem{rikv94_review}
P.~A. Rikvold and B. Gorman,  in {\em Annual Reviews of Computational Physics
  I}, edited by D. Stauffer (World Scientific, Singapore, 1994), p.\ 149.

\bibitem{geilo97}
P.~A. Rikvold, M.~A. Novotny, M. Kolesik, and H.~L. Richards,  in {\em
  Dynamical Properties of Unconventional Magnetic Systems}, edited by A.~T.
  Skjeltorp and D. Sherrington (Kluwer, Dordrecht, 1997).

\bibitem{side96}
S.~W. Sides, R.~A. Ramos, P.~A. Rikvold, and M.~A. Novotny, J.\ Appl.\ Phys.\
  {\bf 79},  6482  (1996).

\bibitem{side97}
S.~W. Sides, R.~A. Ramos, P.~A. Rikvold, and M.~A. Novotny, J.\ Appl.\ Phys.\
  {\bf 81},  5597  (1997).

\bibitem{side98-MMM}
S.~W. Sides, P.~A. Rikvold, and M.~A. Novotny, In press J.\ Appl.\ Phys.\
  Preprint cond-mat/9710244  (1988).

\bibitem{side98-MD}
S.~W. Sides, P.~A. Rikvold, and M.~A. Novotny, In preparation  (1998).

\bibitem{acha94_review}
M. Acharyya and B.~K. Chakrabarti,  in {\em Annual Reviews of Computational
  Physics I}, edited by D. Stauffer (World Scientific, Singapore, 1994), p.\
  107.

\bibitem{acha92-1}
M. Acharyya and B.~K. Chakrabarti, Physica A {\bf 192},  471  (1992).

\bibitem{acha92-2}
M. Acharyya, B.~K. Chakrabarti, and A.~K. Sen, Physica A {\bf 186},  231
  (1992).

\bibitem{acha94}
M. Acharyya and B.~K. Chakrabarti, J.\ Magn.\ Magn.\ Mater.\ {\bf 136},  L29
  (1994).

\bibitem{acha94-2}
M. Acharyya, B.~K. Chakrabarti, and R. Stinchcombe, J.\ Phys.\ A: Math.\ Gen.\
  {\bf 27},  1533  (1994).

\bibitem{lo90}
W. Lo and R.~A. Pelcovits, Phys.\ Rev.\ A {\bf 42},  7471  (1990).

\bibitem{fan95-2}
Z. Fan, Z. Jinxiu, and L. Xiao, Phys.\ Rev.\ E. {\bf 52},  1399  (1995).

\bibitem{acha97}
M. Acharyya, Phys.\ Rev.\ E {\bf 56},  1234  (1997).

\bibitem{acha97-2}
M. Acharyya, Physica A {\bf 235},  469  (1997).

\bibitem{dhar92}
D. Dhar and P.~B. Thomas, J.\ Phys.\ A: Math.\ Gen.\ {\bf 25},  4967  (1992).

\bibitem{thom93}
P.~B. Thomas and D. Dhar, J.\ Phys.\ A: Math.\ Gen.\ {\bf 26},  3973  (1993).

\bibitem{acha95}
M. Acharyya and B.~K. Chakrabarti, Phys.\ Rev.\ B {\bf 52},  6550  (1995).

\bibitem{bind88}
K. Binder and D. Heermann, {\em Monte Carlo Simulation in Statistical Physics}
  (Springer, Berlin, 1988).

\bibitem{mart77}
P.~A. Martin, J.\ Stat.\ Phys.\ {\bf 16},  149  (1977).

\bibitem{jlee95}
J. Lee, M.~A. Novotny, and P.~A. Rikvold, Phys.\ Rev.\ E {\bf 52},  356
  (1995).

\bibitem{jlee95data}
The free-energy data shown in Fig.~\protect\ref{F(m)} were obtained by a
  multicanonical simulation of a 64$\times$64 square-lattice Ising ferromagnet
  at $T = 0.8 T_c$ as part of the research reported in
  Ref.~\protect\cite{jlee95}. This figure is analogous to Fig.~1(a) of that
  paper, which shows the free energy for a 24$\times$24 system at $H$=0  .

\bibitem{lang67}
J.~S. Langer, Ann.\ Phys.\ (N.Y.) {\bf 41},  108  (1967).

\bibitem{lang69}
J.~S. Langer, Ann.\ Phys.\ (N.Y.) {\bf 54},  258  (1969).

\bibitem{gnw80}
N.~J. G{\"u}nther, D.~A. Nicole, and D.~J. Wallace, J.\ Phys.\ A {\bf 13},
  1755  (1980).

\bibitem{ccga94}
C.~C.~A. G\"{u}nther, P.~A. Rikvold, and M.~A. Novotny, Physica A {\bf 212},
  194  (1994).

\bibitem{ramo97}
R.~A. Ramos, S.~W. Sides, P.~A. Rikvold, and M.~A. Novotny, In preparation  .

\bibitem{cox65}
D. Cox and H. Miller, {\em The Theory of Stochastic Processes} (Methuen,
  London, 1965).

\bibitem{zhou90-2}
T. Zhou, F. Moss, and P. Jung, Phys.\ Rev.\ A {\bf 42},  3161  (1990).

\bibitem{gunt83}
J. Gunton and M. Droz, {\em Introduction to the Theory of Metastable and
  Unstable States} (Springer, Berlin, 1983).

\bibitem{lifs62}
I. Lifshitz, Sov.\ Phys.\ JETP {\bf 15},  939  (1962).

\bibitem{alle79}
S. Allen and J. Cahn, Acta Metall.\ {\bf 27},  1085  (1979).

\bibitem{seki86}
K. Sekimoto, Physica {\bf 135A},  328  (1986).

\bibitem{shne94-2}
V. Shneidman and P. H\"{a}nggi, Phys.\ Rev.\ E. {\bf 49},  641  (1994).

\bibitem{jung93}
P. Jung, Phys.\ Rep.\ {\bf 234},  175  (1993).

\bibitem{gamm95}
L. Gammaitoni, F. Marchesoni, and S. Santucci, Phys.\ Rev.\ Lett. {\bf 74},
  1052  (1995).

\bibitem{wern97-1}
W. Wernsdorfer {\it et~al.}, Phys.\ Rev.\ Lett. {\bf 78},  1791  (1997).

\bibitem{pres92}
W.~H. Press, S.~A. Teukolsky, W.~T. Vetterling, and B.~P. Flannery, {\em
  Numerical Recipes in Fortran: The Art of Scientic Computing}, 2nd  ed.
  (Cambridge University Press, Boston, 1992).

\bibitem{shne94-1}
V. Shneidman, P. Jung, and P. H\"{a}nggi, Europhys.\ Lett. {\bf 26},  571
  (1994).

\bibitem{zhou90}
T. Zhou and F. Moss, Phys.\ Rev.\ A {\bf 41},  4255  (1990).

\bibitem{side98_thesis}
S.~W. Sides, Ph.D. thesis, In preparation  (1998).

\bibitem{abra70}
M. Abramowitz and I.~A. Stegun, {\em Handbook of Mathematical Functions}
  (National Bureau of Standards, Washington, D.C., 1970), p. 260.

\bibitem{maha94}
M.~C. Mahato and S.~R. Shenoy, Phys.\ Rev.\ E {\bf 50},  2503  (1994).

\bibitem{maha97}
M.~C. Mahato and A.~M. Jayannavar, Phys.\ Rev.\ E {\bf 55},  6266  (1997).

\bibitem{acha98}
M. Acharyya, Preprint cond-mat/9712309  (1998).

\end{thebibliography}

 \begin{table}
 \begin{center}
 \normalsize
 \begin{tabular}{|| l | r || l | r || l | r ||}
  \multicolumn{2}{||c||}{Parameters} 
  & \multicolumn{2}{c||}{Constants (theory)} & 
  \multicolumn{2}{c||}{Constants (simulation)} \\
  \hline
  $H_{0}$  &   $0.1J$    &  $\Xi_{0}(T)$  & $0.506192 \ J$
                                                            
&   $\Omega$   &  $3.15255$  \\
  \hline
  $L$      &   $64$       &  $K$        &     $3$ (exact)   &   $\nu$      
           & $(0.465 \pm 0.014) \ J^{-1}{\rm MCSS}^{-1}$ \\
  \hline
  $T$      &   $0.8T_{c}$ & \multicolumn{2}{c||}{}          
           &   $\left < \tau \right > $     &  $2058$ MCSS \\
  \hline
  \multicolumn{2}{||c||}{}& \multicolumn{2}{c||}{}          
  & $H_{\rm DSP}$& $(0.11 \pm 0.005) \ J$  \\
  \hline
  \multicolumn{2}{||c||}{}& \multicolumn{2}{c||}{}          
  & $r$          & $0.672$   \\
 \end{tabular}
 \end{center}
 \caption{
  \label{table_constants}
  Parameters and constants used in this work.
  The values of the parameters $H_{0}$, $L$, and $T$ have
  been selected
  such that switching occurs via the single-droplet mechanism,
  while the maximum nucleation rate is not too low to obtain reasonable
  simulation statistics.
  The constants $\Xi_{0}(T)$ and $K$
  are calculated from droplet theory 
  \protect\cite{lang67,lang69,gnw80,ccga94}
  for two-dimensional Ising systems.
  The constants
  $\left < \tau \right >$ and $r$ are measured from 
  field-reversal MC simulations
  with the Glauber dynamic
  (using the parameters listed above).
  The constants
  $\Omega$ \protect\cite{ccga94}
  and $\nu$ \protect\cite{ramo97}
  have been measured in other work (for clarity, we do not explicitly show
   the temperature dependence of these quantities in the table
   or elsewhere in the paper).
  The value for $H_{\rm DSP}$ is taken from
  Fig. 11 of Ref.~\protect\cite{jlee95}.}
 \end{table}


\newpage

\newpage
\begin{figure}[tbp]
\vspace*{3.5in}
\includegraphics{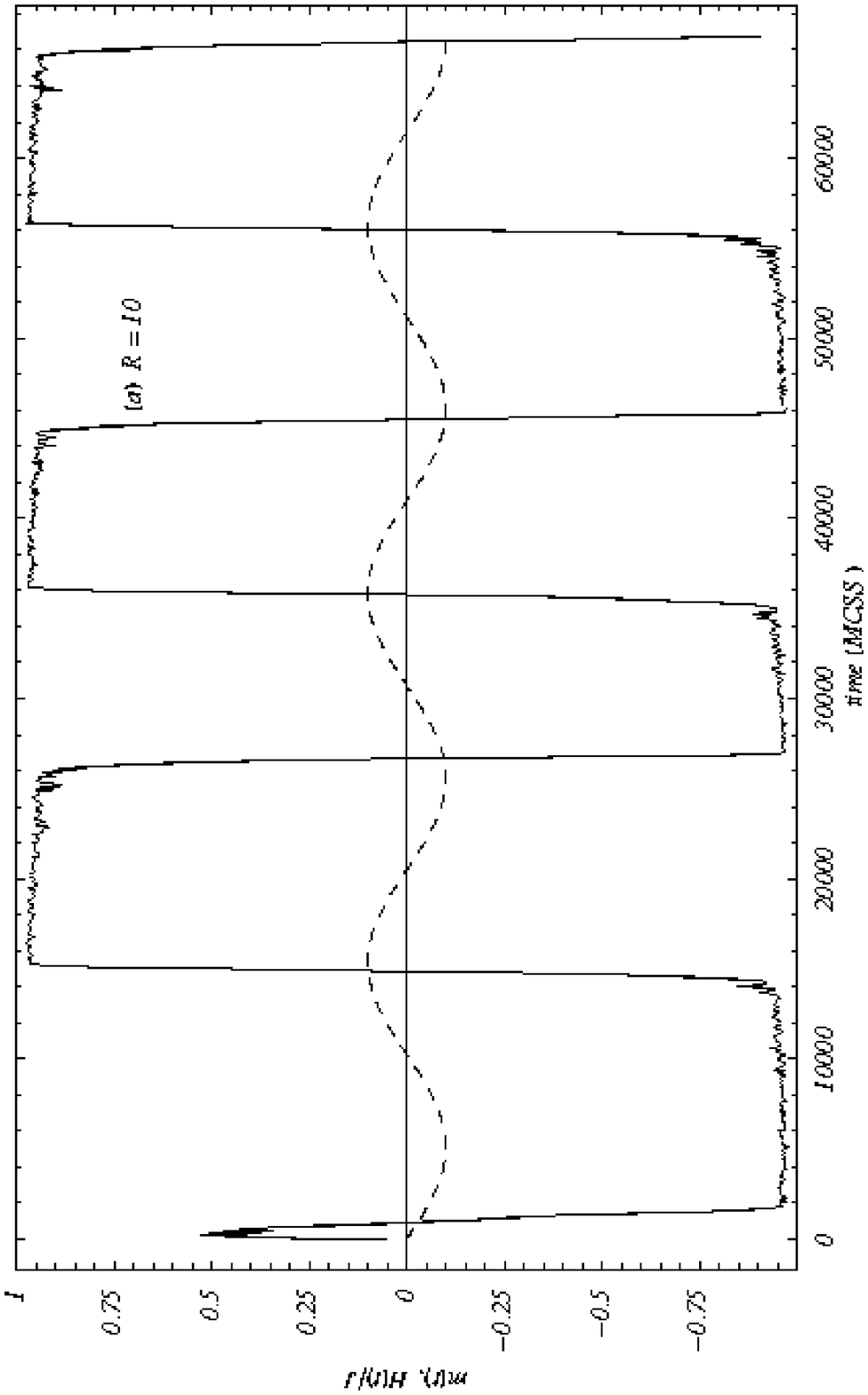}
\bigskip
\vspace*{3.75in}
\includegraphics{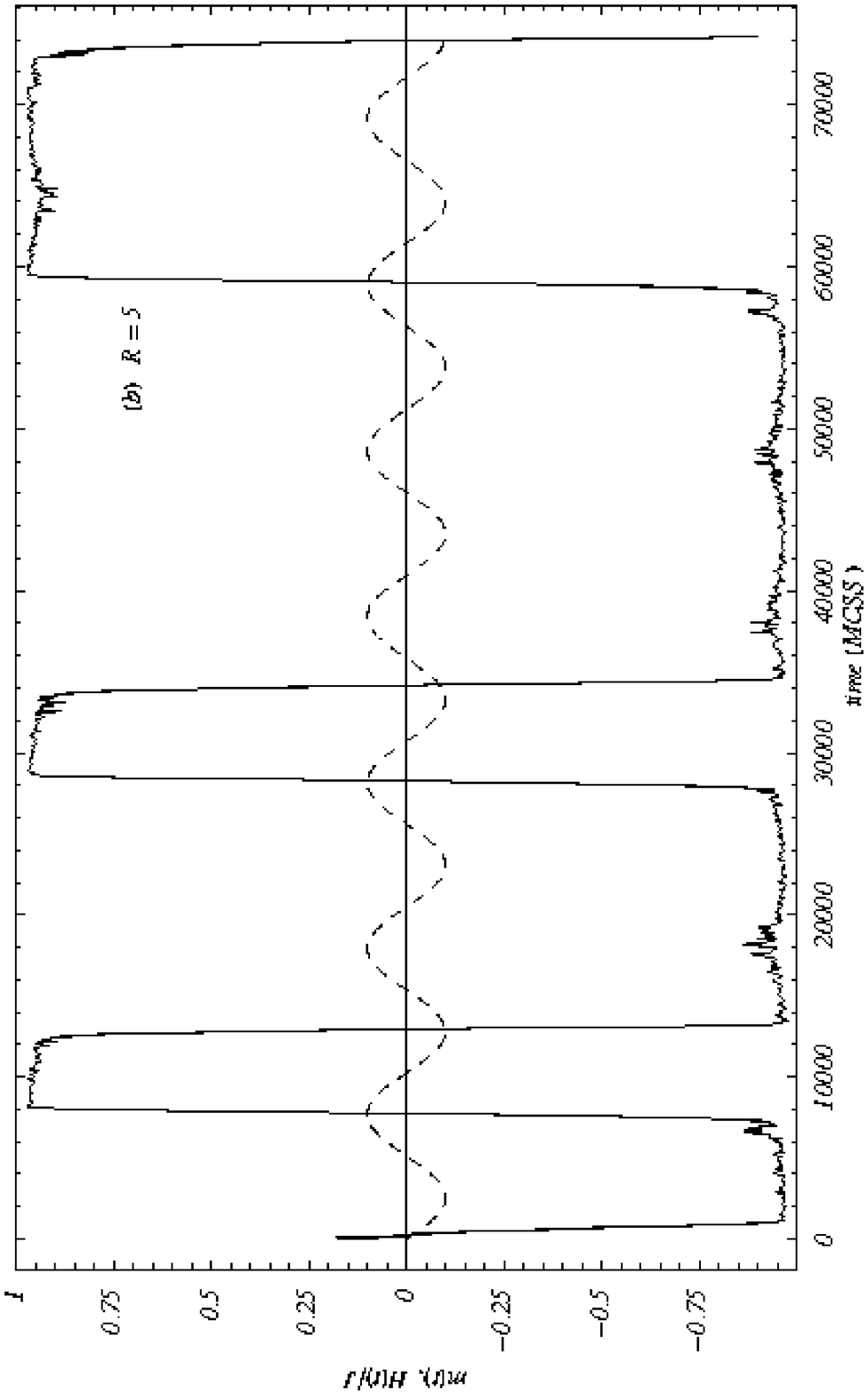}
\bigskip
\end{figure}
\newpage
\begin{figure}[tbp]
\vspace*{3.5in}
\includegraphics{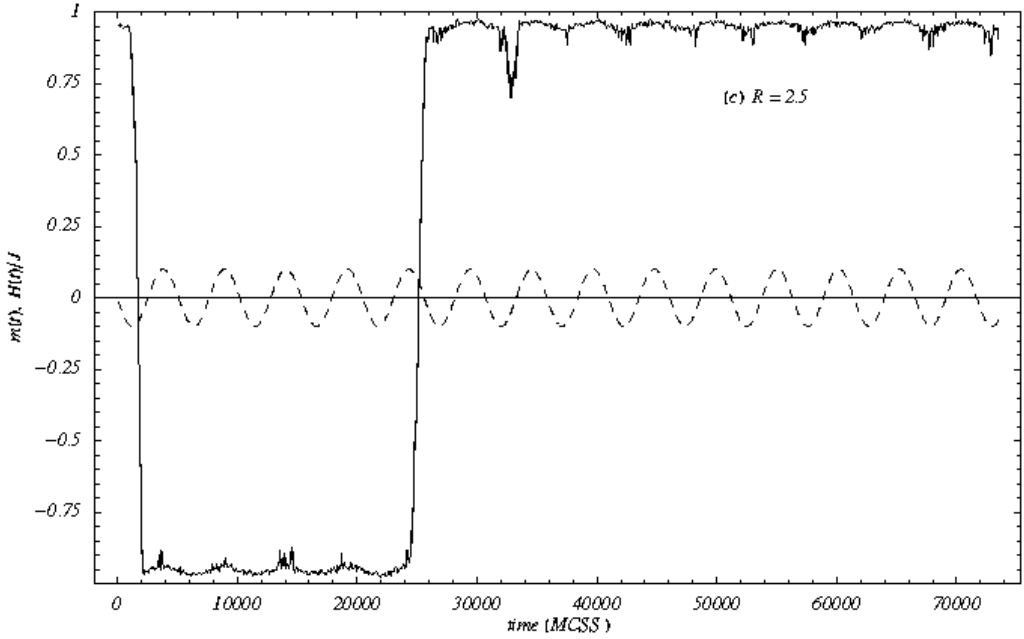}
\bigskip
\caption{ \label{mtSD}
Short initial segments of the
magnetization time series $m(t)$ (solid line)
and the external field $H(t)$ (dashed line)
vs.\ time $t$ in the SD region, for $T$=$0.8T_{c}$,
$d$=$2$, $L$=$64$, and $H_{0}=0.1J$.
The total length of the time series is
approximately $16.9 \times 10^{6}$ MCSS.
For these parameter values the average lifetime in static field
is $\left < \tau (H_{0}) \right > \approx 2058$ MCSS.
The time series are shown for the scaled field periods
(a) $R$=$10$,
(b) $R$=$5$, and
(c) $R$=$2.5$}
\end{figure}

\begin{figure}[tbp]
\vspace*{3.5in}
\includegraphics{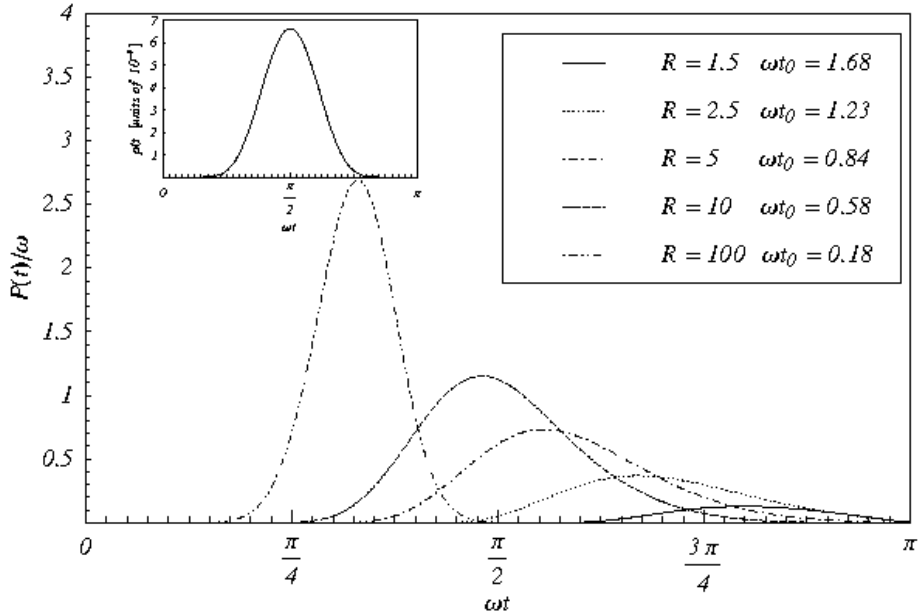}
\bigskip
\caption{\label{fig_rho}
The switching probability density, $P(t)/\omega$ vs.\ $\omega t$,
for five values of the period of the external field,
$R$=$1.5$, $R$=$2.5$, $R$=$5$, $R$=$10$, and $R$=$100$.
The plots are obtained from a numerical evaluation of
Eq.~(\protect\ref{eq_P(t)}).
The inset shows the decay rate, $\rho (t)$ vs.\ $\omega t$.
The time $t_{0}$ is the earliest time during a period,
for which $P(t)$ is non-zero.
Even though the decay rate always has a maximum when the phase
equals $\pi/2$,
the value of the phase for which $P(t)/\omega$ is maximum depends on the
frequency.}
\end{figure}

\newpage
\begin{figure}[tbp]
\vspace*{3.5in}
\includegraphics{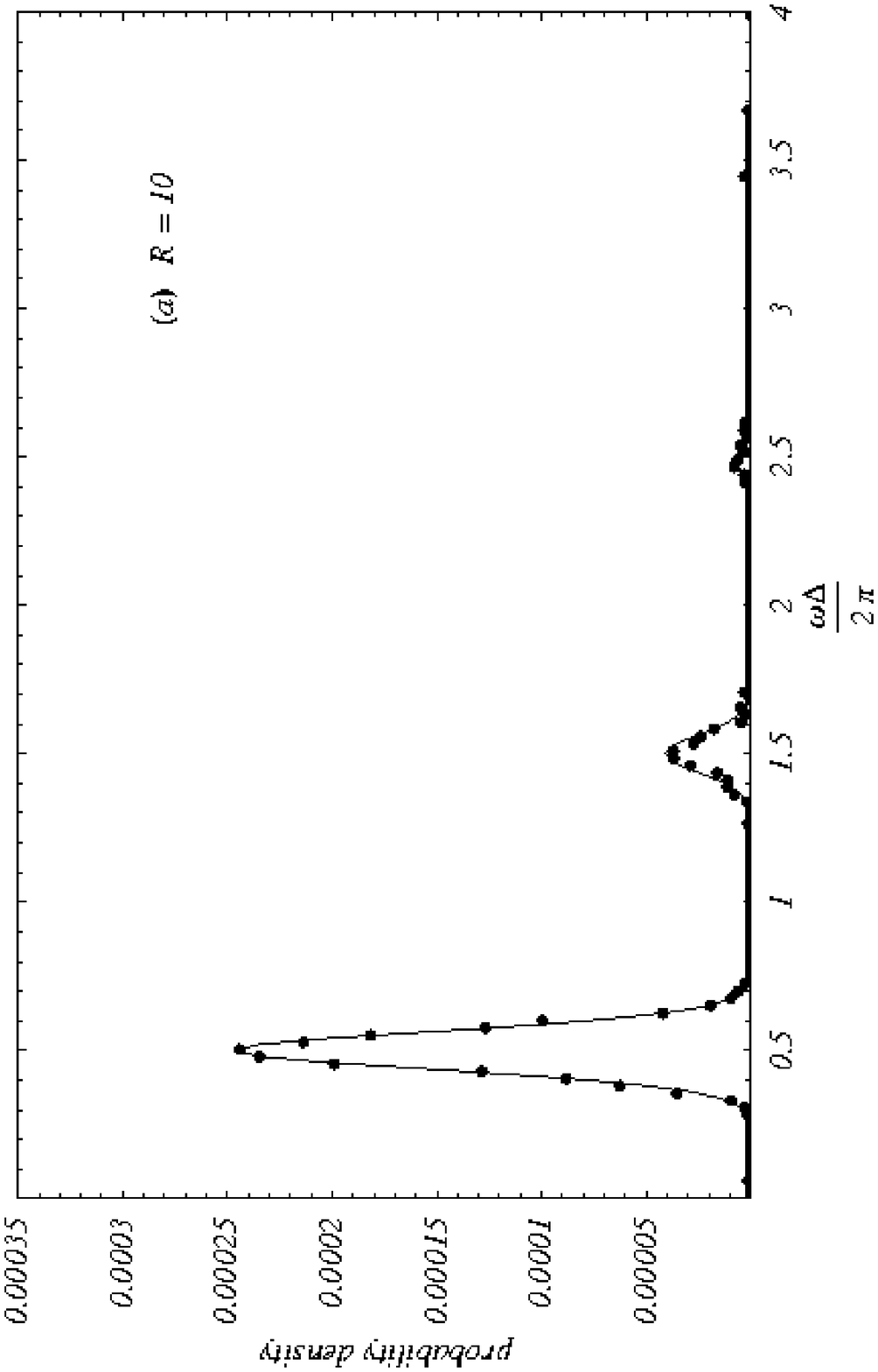}
\vspace*{3.75in}
\includegraphics{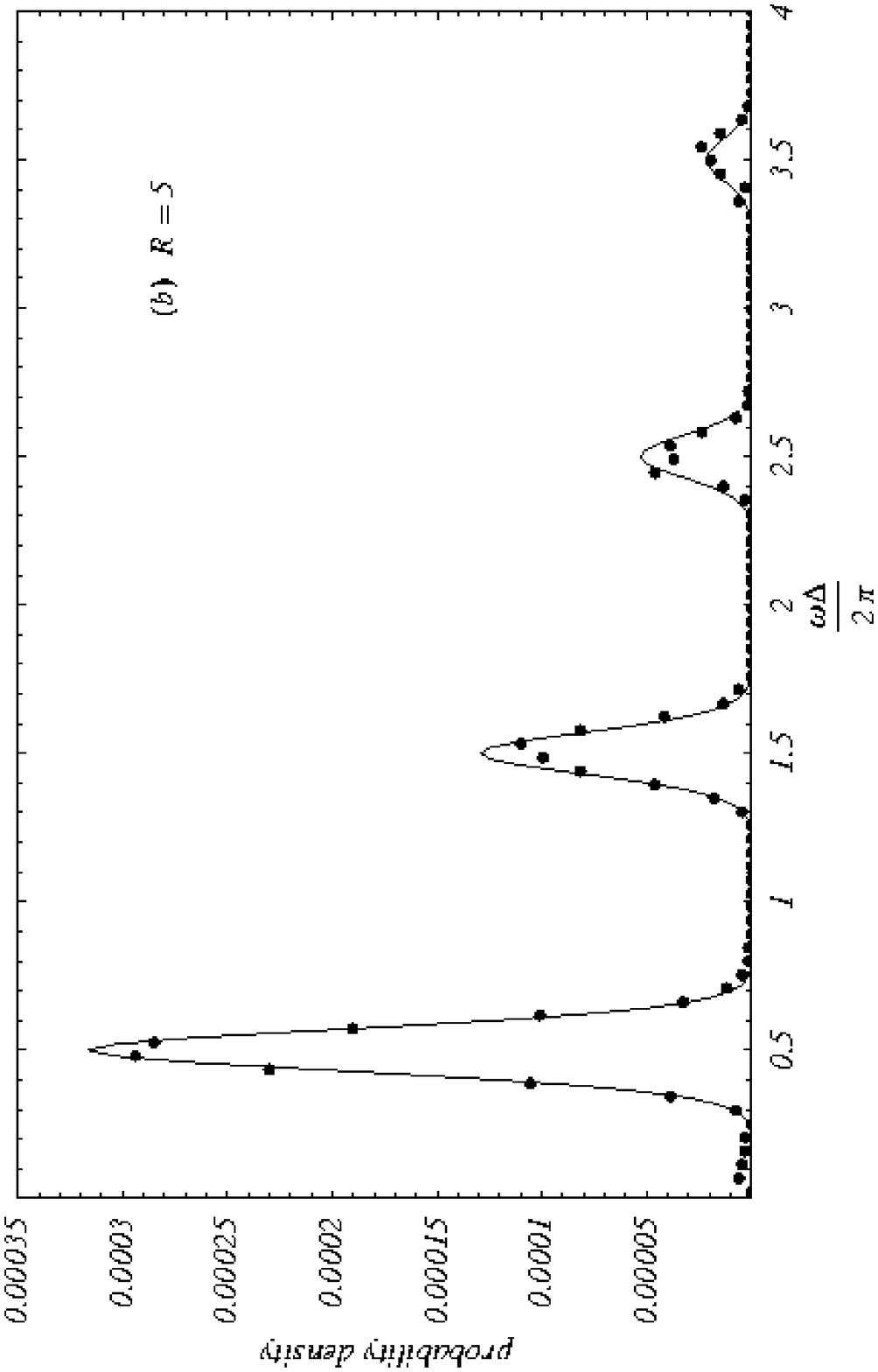}
\bigskip
\end{figure}
\newpage
\begin{figure}[tbp]
\vspace*{3.5in}
\includegraphics{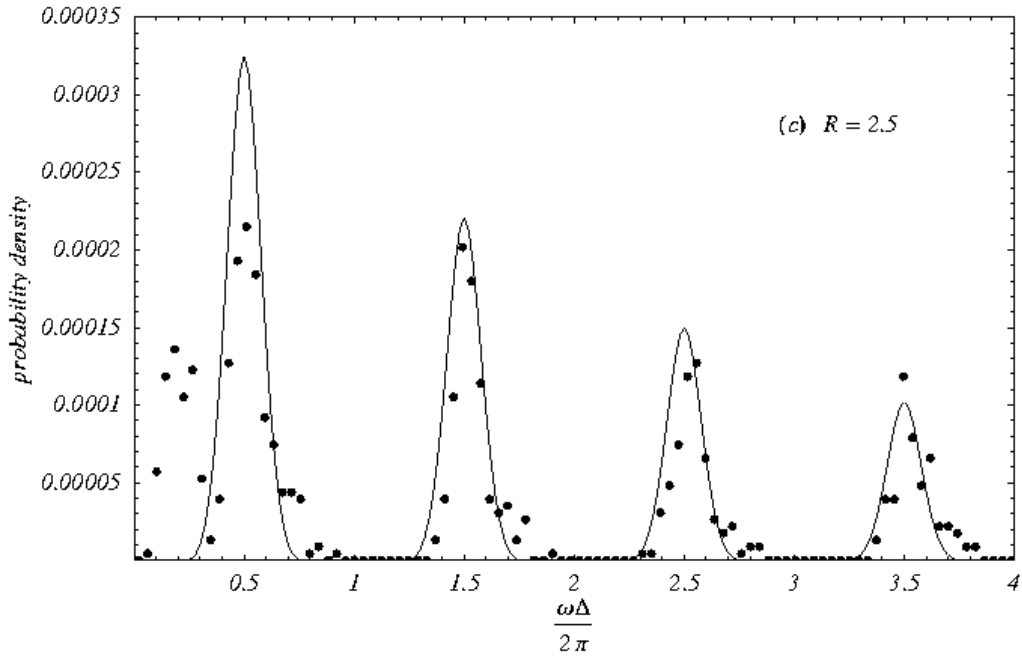}
\bigskip
\caption{\label{rtdSD}
Residence-time distributions (RTDs).
The time axis is scaled by the period $2 \pi/\omega$ of the
external field for each value of $R$, so that the
peaks are centered around odd half-integer multiples.
The RTDs are shown for
(a) $R$=$10$     (150 bins, 1239 events),
(b) $R$=$5$      (250 bins, 1439 events), and
(c) $R$=$2.5$    (500 bins, 1089 events).
The filled circles are obtained from the MC simulations.
The solid curves represent the theoretical calculation presented in
Appendix \protect\ref{appendix_residencetime_dis}.
}
\end{figure}

\begin{figure}[tbp]
\vspace*{3.5in}
\includegraphics{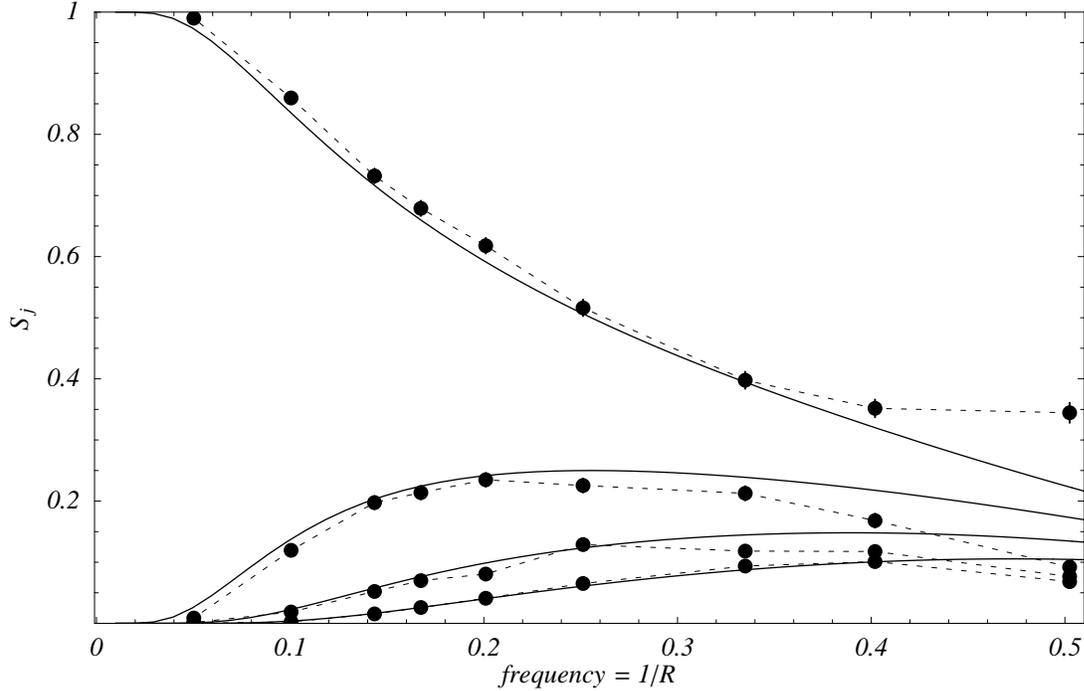}
\bigskip
\caption{\label{rtdSDpeak}
Peak strengths in the RTDs vs.\ scaled frequency $1/R$.
The different curves, from top to bottom, correspond to
$S_{1}$, $S_{2}$, $S_{3}$, and $S_{4}$.
The statistical errors are everywhere on the order of the
symbol size or less.
The solid curves, obtained from Eq.~(\protect\ref{eq_Somega}),
result from the same parameter-free calculation as the RTDs.
}
\end{figure}

\newpage
\begin{figure}[tbp]
\vspace*{3.5in}
\includegraphics{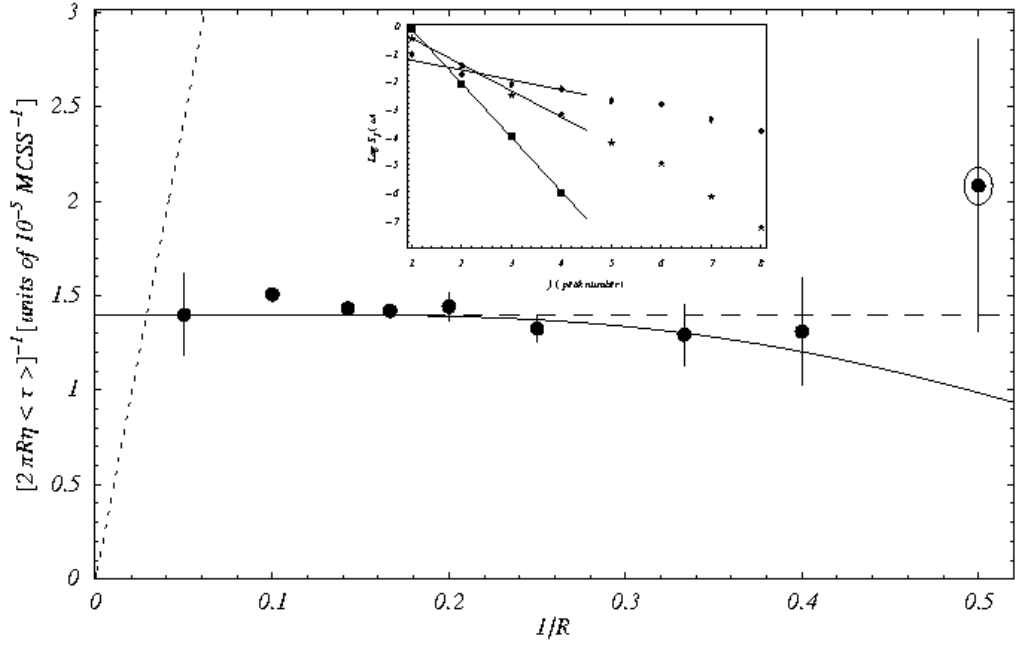}
\bigskip
\caption{\label{etaSD}
The inverse characteristic time of the RTDs,
$\left[2 \pi R \eta \left < \tau \right > \right]^{-1}$,
vs.\ the scaled frequency, $1/R$.
The inverse characteristic time
is given in units of $10^{-5} {\rm MCSS}^{-1}$.
The solid dots are calculated from MC data for the peak strengths
of the RTDs.
The inset shows examples of $\ln S_{j}$ vs.\
the peak number $j$, along
with their linear least-squares fit lines.
The three frequencies shown in the inset
are $1/R$=$0.4$, $0.2$, and $0.1$,
denoted by diamonds, stars, and squares respectively.
The error bars on the data
are calculated from the standard deviation
in the slopes of the fitted lines.
The solid curve is obtained from the
full numerical calculation of Eq.~(\protect\ref{eq_eta_theory}).
The horizontal long-dashed line
results from a low-frequency approximation obtained by setting
$t_{\rm g}$=$0$.
This horizontal line is located at a value of
$\left[ 2 \pi R \eta \left < \tau \right > \right ]^{-1}$
=$1.4 \times 10^{-5} {\rm MCSS}^{-1}$
obtained by numerical
integration of Eq.~(\protect\ref{eq_chartime_analytic1}).
The short-dashed line represents the frequency,
$\left < \tau \right >/R$,
which has been scaled to have the
units of ${\rm MCSS}^{-1}$.
The open oval around the point for $1/R$=$0.5$ is a
reminder of the poor statistics
and large systematic error in the RTD for this frequency.
}
\end{figure}

\newpage
\begin{figure}[tbp]
\vspace*{3.5in}
\includegraphics{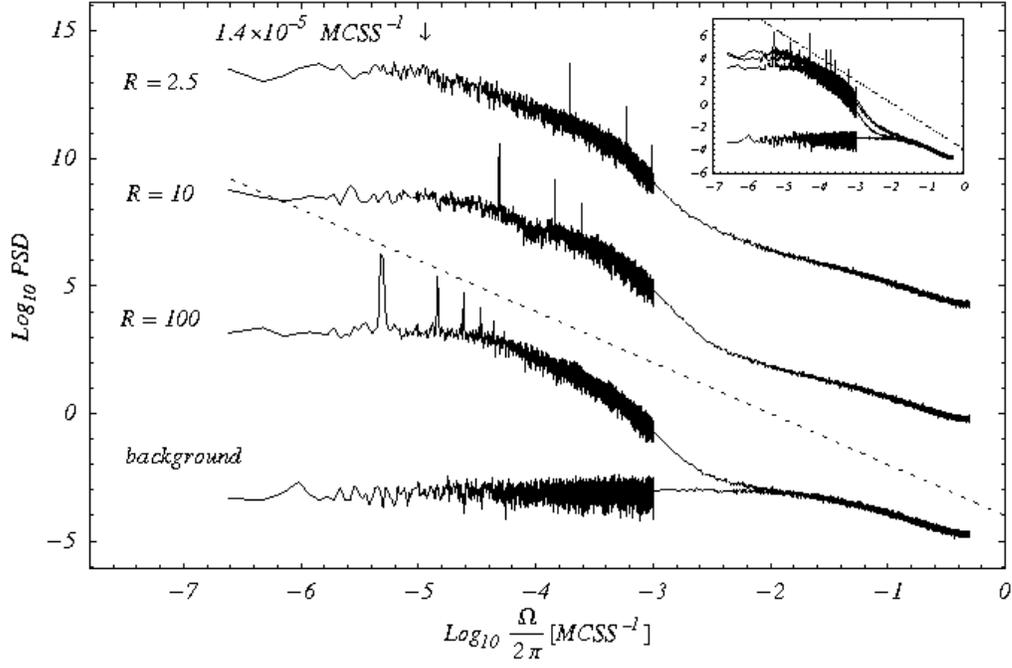}
\bigskip
\caption{\label{psdSD}
Power spectral densities (PSDs).
Spectra are shown for three different frequencies of the
external field, and are plotted with an arbitrary
offset for clarity.
The inset shows the same spectra without the offset
to illustrate how all three PSDs fall onto the
thermal noise background at high frequencies.
In addition to the change in the amount of smoothing,
the right-hand section of each spectrum contains only one data point
out of every 25 to facilitate plotting.
The magnetization is sampled every 1.0 MCSS, so the Nyquist
frequency is 0.5 MCSS$^{-1}$.
The lowest frequency that can be resolved is
$2.38 \times 10^{-7} \ {\rm MCSS}^{-1}$.
The dashed line with slope $-2$
is a guide to the eye.
The arrow indicates a frequency of
$1.4 \times 10^{-5} {\rm MCSS}^{-1}$ in the PSD.
This frequency value is the half-width of the spectrum
predicted in Sec.~\protect\ref{subsec_chartime} for
low frequencies of the external field.
}
\end{figure}

\newpage
\begin{figure}[tbp]
\vspace*{3.5in}
\includegraphics{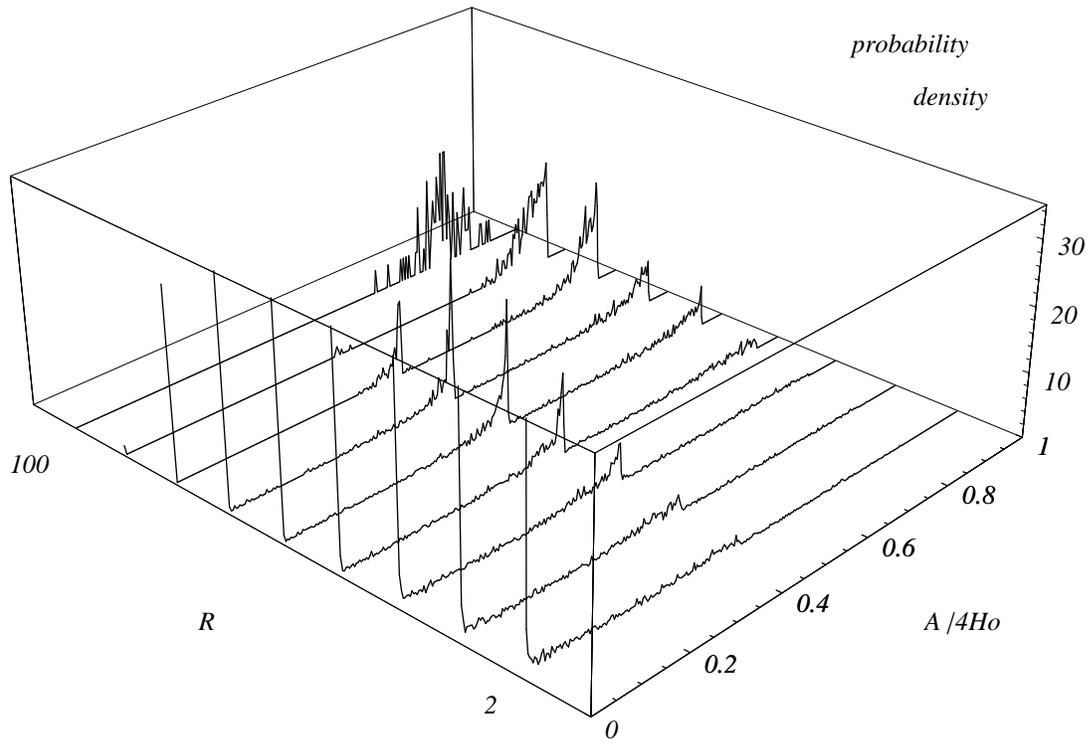}
\bigskip
\caption{\label{aSD} 
Probability densities for the hysteresis-loop area,
$A$=$- \oint m(H) \ dH$.
The values of $R$ shown here are
$R$=$2$,
$2.5$,
$3$,
$4$,
$5$,
$6$,
$10$,
$20$,
and $100$.}
\end{figure}

\begin{figure}[tbp]
\vspace*{3.65in}
\includegraphics{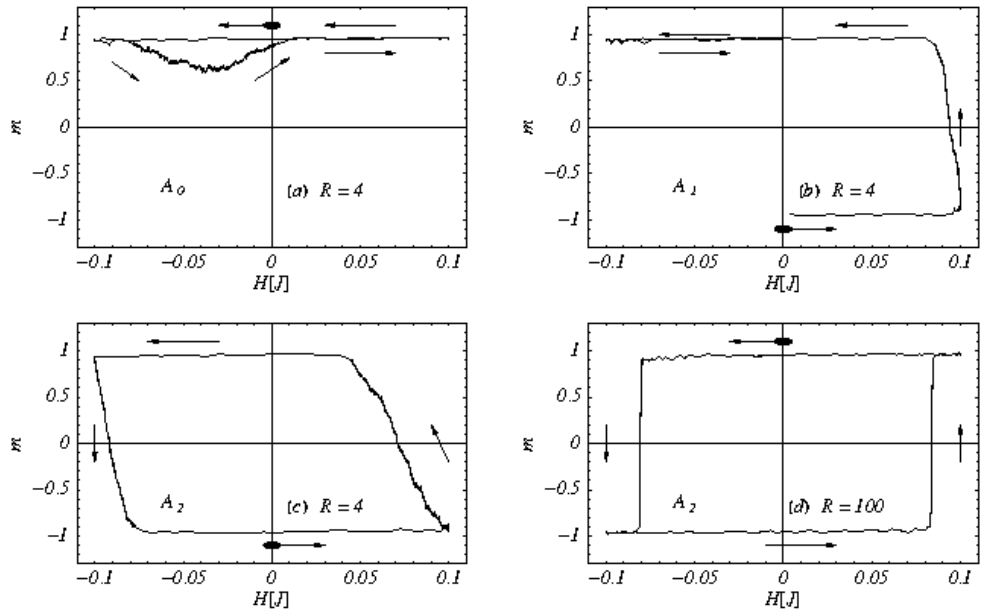}
\bigskip
\caption{\label{aSD_loops}
Representative hysteresis loops
corresponding to different numbers of switching events during
one period of the applied field.
Each loop in panels (a)-(c) consists of data from a {\it single} 
period in a time series with $R$=$4$.
For $A_{0}$,
no magnetization reversal occurs.
This example shows a ``spike'' in
the magnetization during one half of the period.
For $A_{1}$,
one magnetization reversal occurs during the period.
For $A_{2}$ in panels (c) and (d),
two magnetization reversals occur.
Panel (d) shows a hysteresis loop of type $A_{2}$
from a very low-frequency time series with $R$=$100$.
Note: the arrow tails with filled dots indicate
the start of the field cycle.}
\end{figure}

\newpage
\begin{figure}[tbp]
\vspace*{3.5in}
\includegraphics{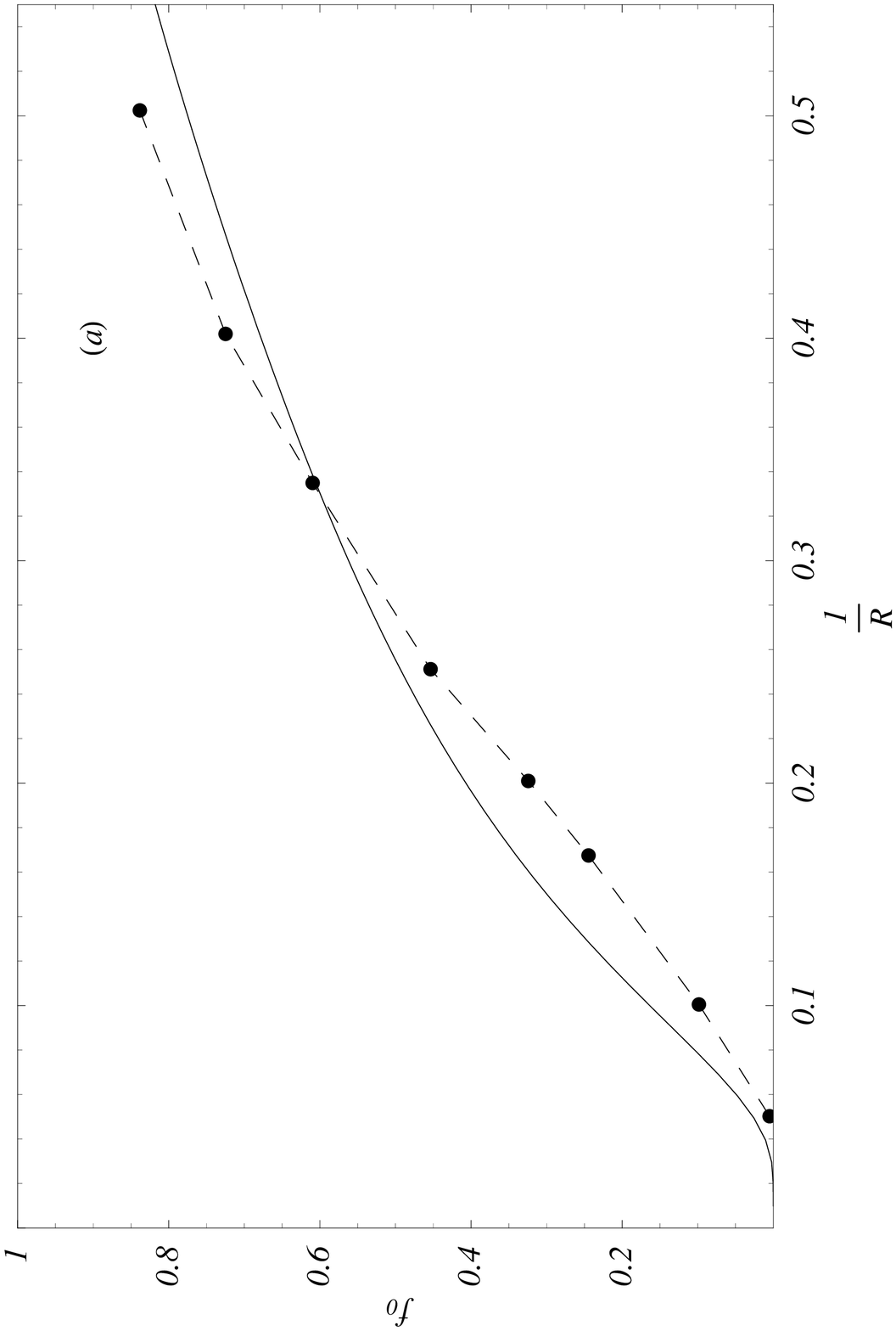}
\vspace*{3.75in}
\includegraphics{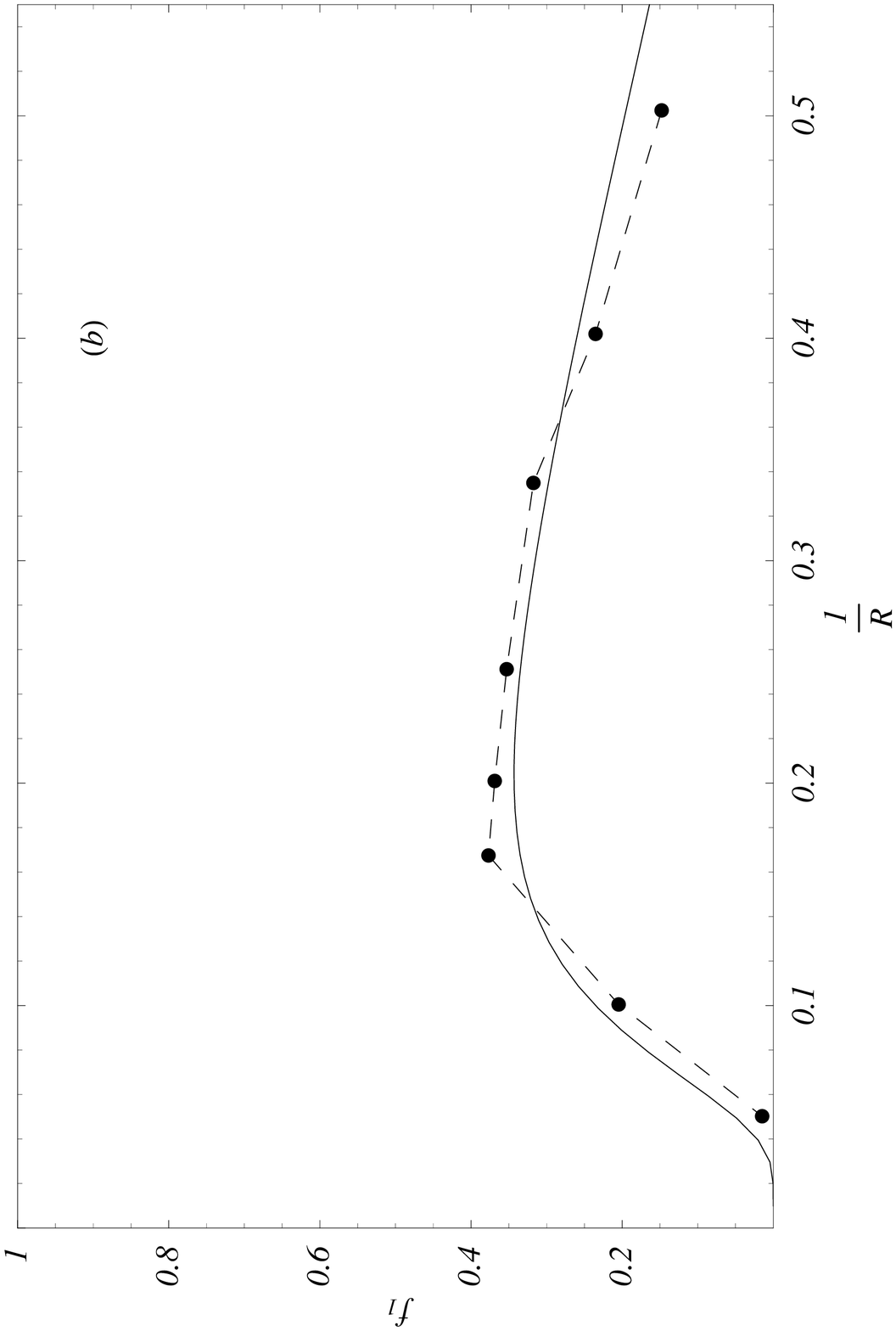}
\bigskip
\end{figure}
\newpage
\begin{figure}[tbp]
\vspace*{3.5in}
\includegraphics{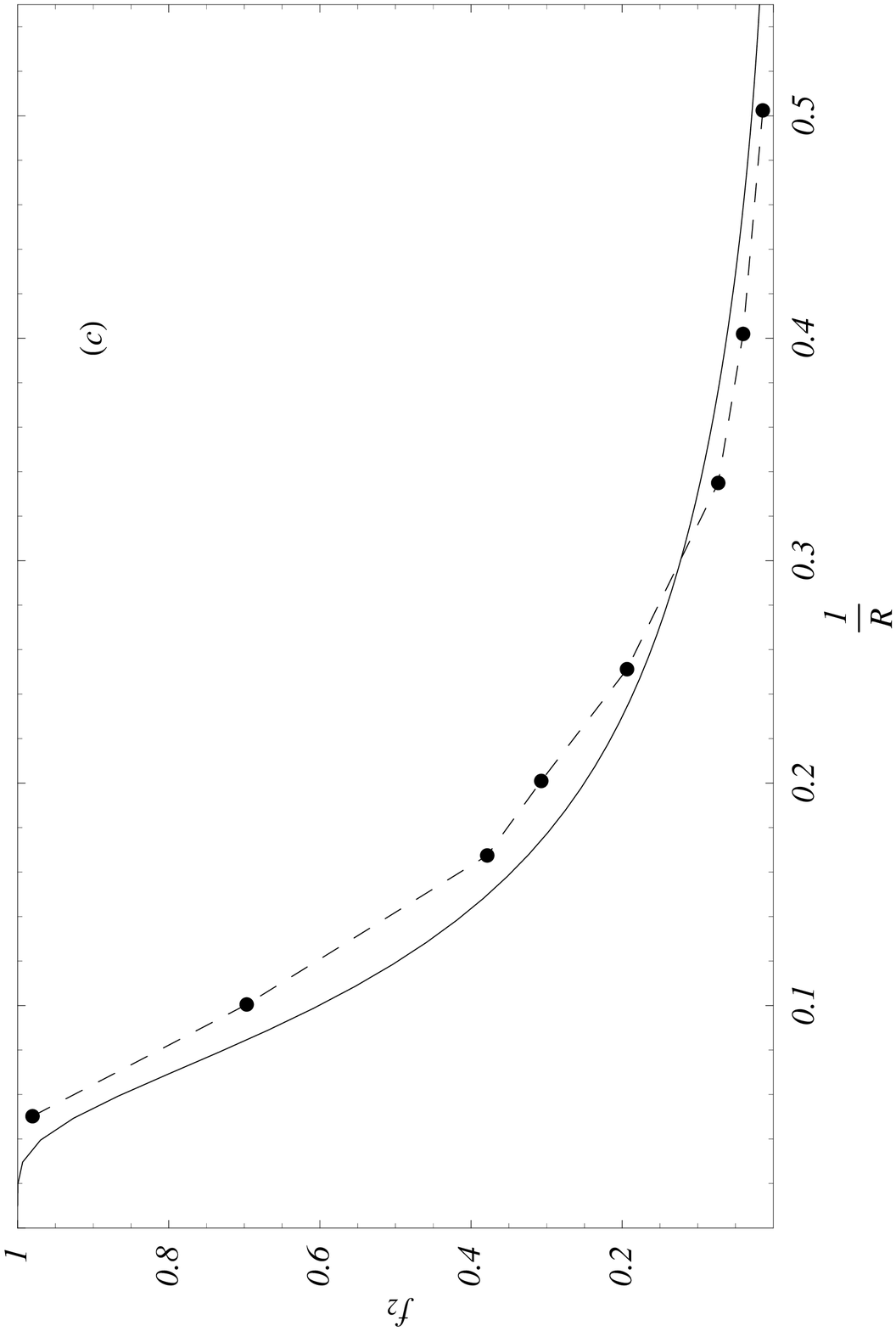}
\vspace*{3.75in}
\includegraphics{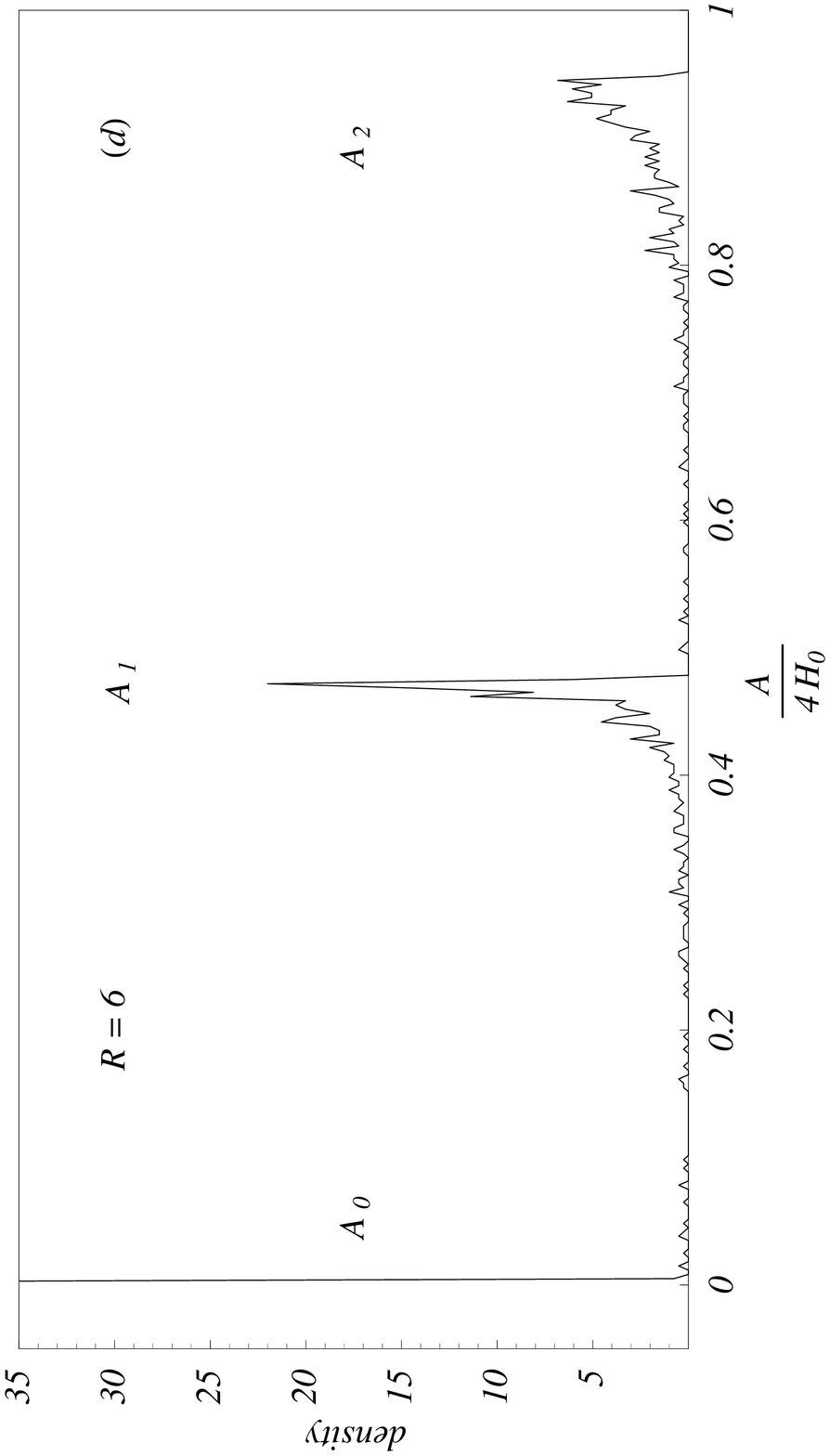}
\bigskip
\caption{\label{aSD_fracs} 
Frequency dependence of the fraction of events
contained in the peaks
(a) $A_{0}$,
(b) $A_{1}$,
and (c) $A_{2}$.
The solid dots are measured directly from the data in the
loop-area distributions, and the dashed lines are guides to the eye.
The solid curve in each plot is calculated using
Eqs.~(\protect\ref{eq_fractions2}a-c)
with $P_{\rm not}(\omega)$ obtained from a numerical evaluation
of Eq.~(\protect\ref{eq_Pnot}).
(d) The loop-area distribution for $R$=$6$ is shown to illustrate
the peak sizes in the trimodal distribution.
}
\end{figure}

\newpage
\begin{figure}[tbp]
\vspace*{3.5in}
\includegraphics{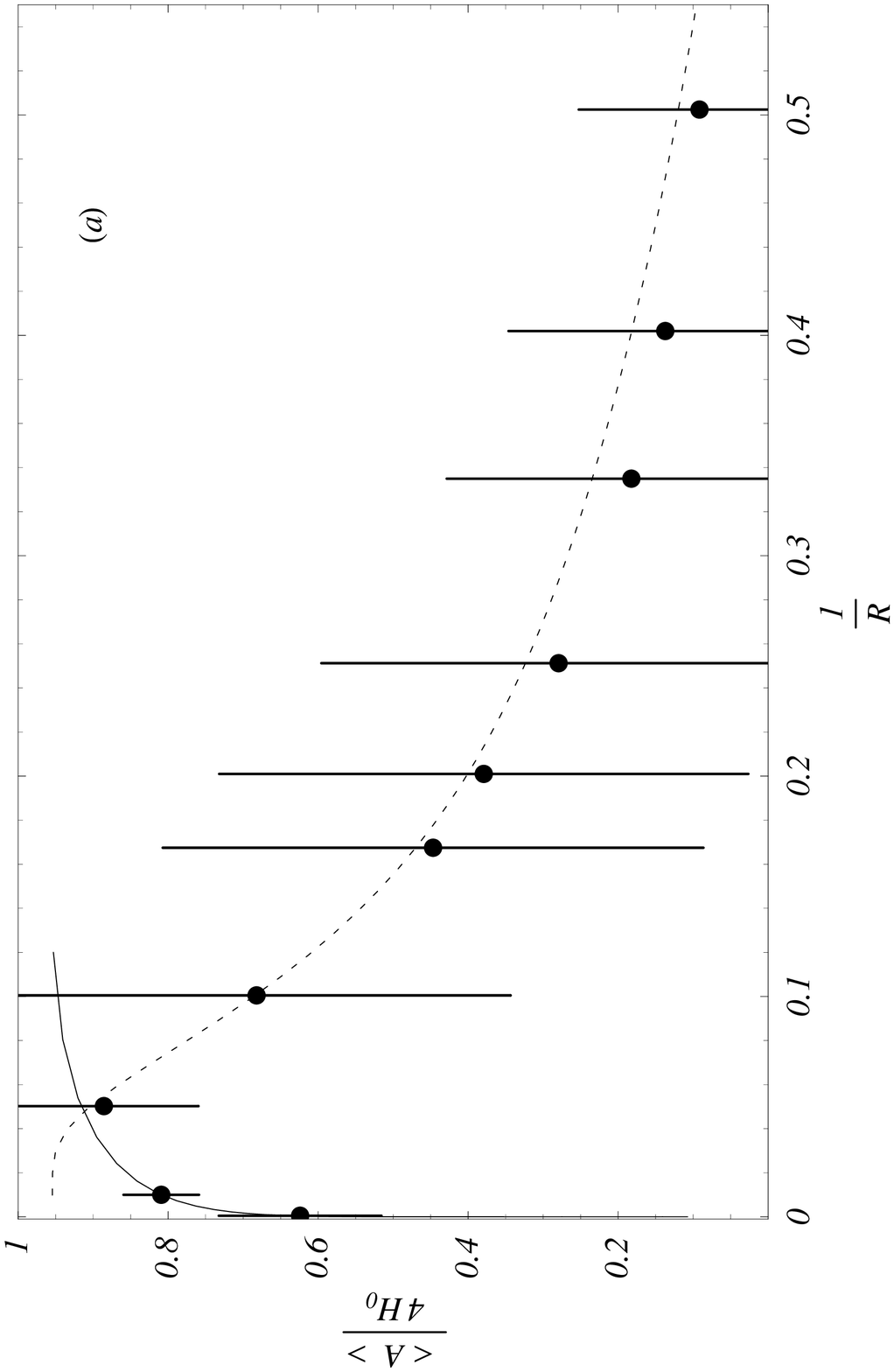}
\bigskip
\end{figure}
\newpage
\begin{figure}[tbp]
\vspace*{3.5in}
\includegraphics{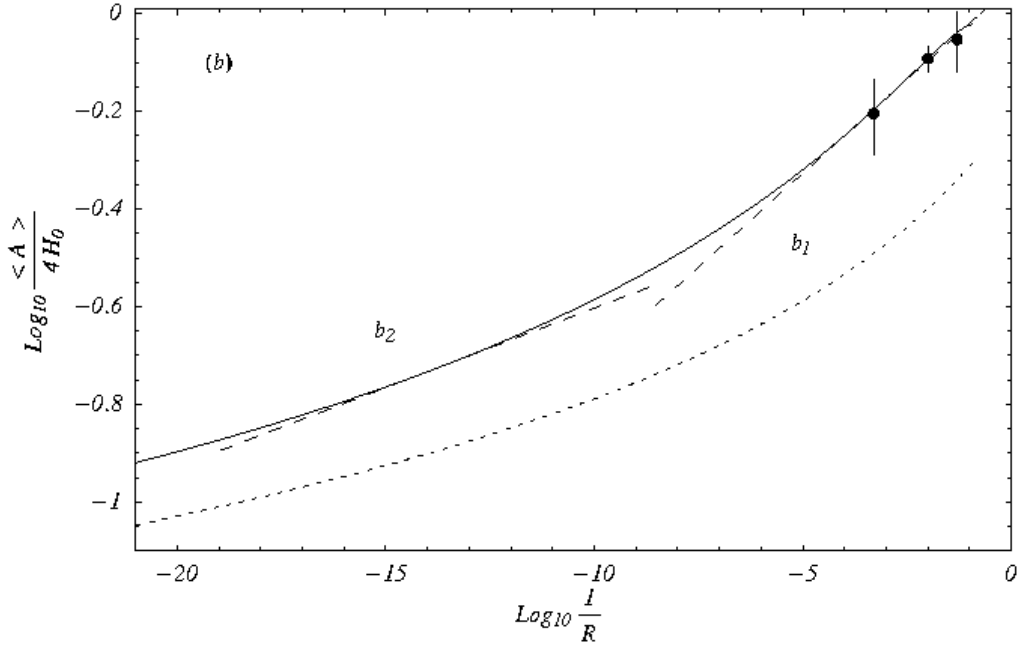}
\bigskip
\caption{\label{aSD_stats}
(a).
Mean and standard deviation of the loop-area
distributions shown vs.\ the scaled frequency, $1/R$.
The solid dots are the means of the distributions shown in
Fig.~\protect\ref{aSD}.
The vertical bars are {\it not} error bars, but
give the standard deviations of those distributions.
The large standard deviations for $1/R \protect\gtrsim 0.1$
indicate the trimodal
distribution of the loop areas, typical of the SD region.
For the points at the two lowest frequencies the
magnetization switches nearly every
half-period, giving a loop-area distribution which is close to
unimodal.
The solid and dotted curves come from two separate theoretical
calculations.
The solid curve results from numerical solution of an analytic
expression for the switching field that assumes switching occurs
during every half-period.
The dotted curve comes from a calculation which uses the same
values of $P_{\rm not}(\omega)$ used for the theoretical curve
in Fig.~\protect\ref{aSD_fracs}.
(b).
Log-log plot for the very lowest frequencies in panel (a).
The solid curve is the same as the solid curve in panel (a).
The long-dashed line segments represent linear least-squares fits
to different portions of the numerical solution data,
each covering almost four decades in frequency.
The data that yield the effective exponent
$b_{1}$=$0.077$, are centered around $\log_{10} (1/R)$=$-3.35$;
those that yield $b_{2}$=$0.033$ are centered
around $\log_{10} (1/R)$=$-13.78$.
The solid dots are MC simulation data.
The short-dashed curve represents Eq.~(\protect\ref{eq_SDloopanalytic})
with $d$=$2$ and $C$=$0.101 J^{-1} {\rm MCSS}$.
}
\end{figure}

\newpage
\begin{figure}[tbp]
\vspace*{3.5in}
\includegraphics{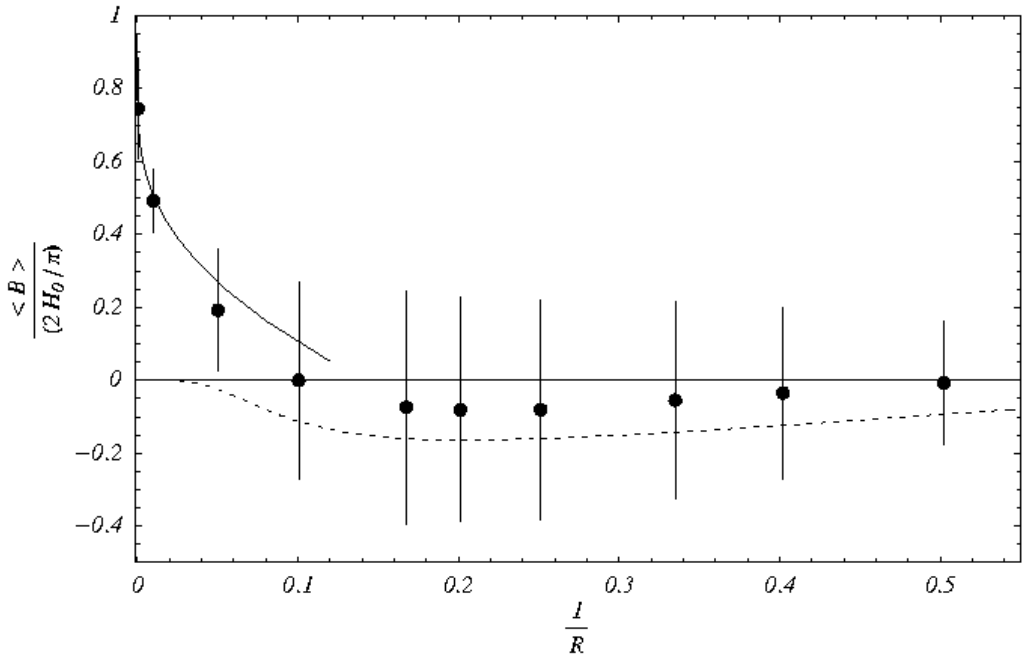}
\bigskip
\caption{\label{bSD} 
Mean and standard deviation of the correlation, $B$
shown vs.\ the scaled frequency, $1/R$.
The solid dots are the means of the correlation distributions.
The vertical bars are {\it not} error bars, but
give the standard deviations of those distributions.
The solid and dotted curves are obtained from similar
calculations as those for the loop areas in
Fig.~\protect\ref{aSD_stats}.
}
\end{figure}

\begin{figure}[tbp]
\vspace*{3.95in}
\includegraphics{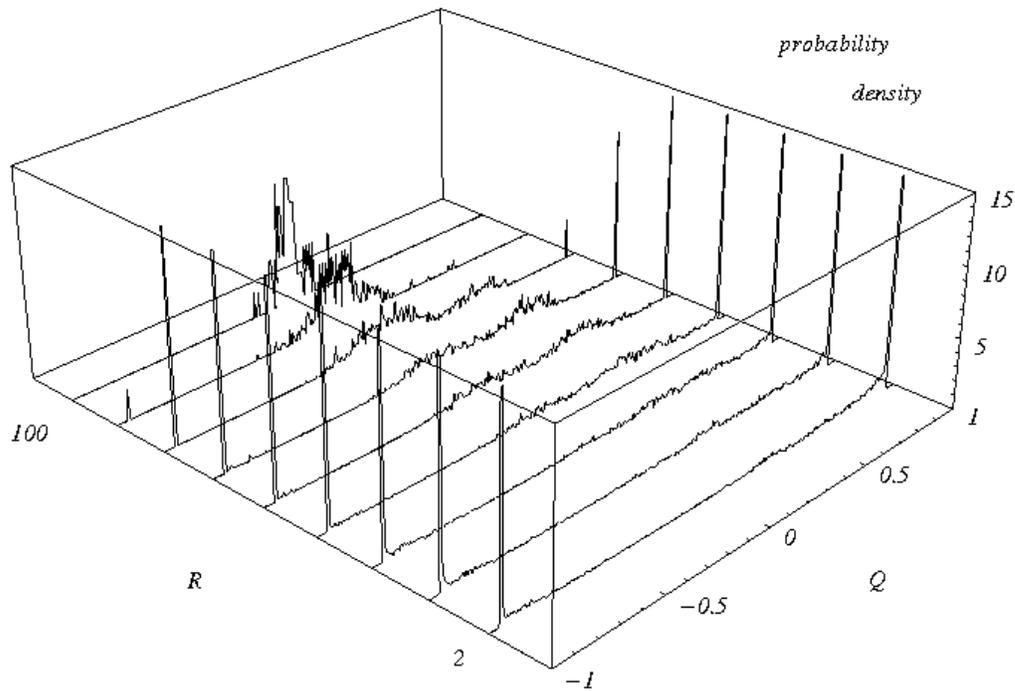}
\bigskip
\caption{\label{qSD} 
Probability densities for
the period-averaged magnetization,
$Q$.
The values of $R$ shown are
$R$=$2$,
$2.5$,
$3$,
$4$,
$5$,
$6$,
$10$,
$20$,
and $100$.}
\end{figure}

\newpage
\begin{figure}[tbp]
\vspace*{3.5in}
\includegraphics{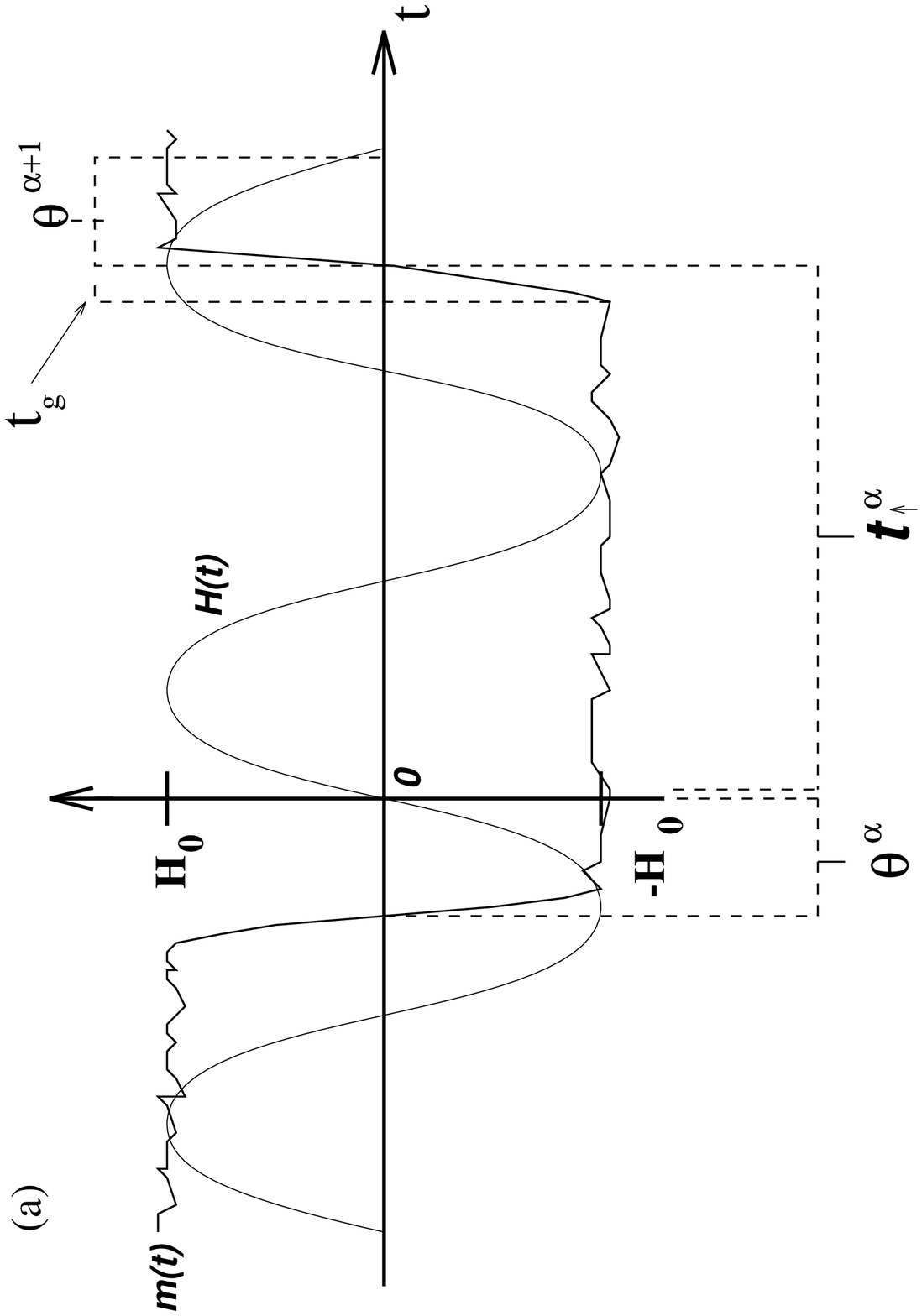}
\vspace*{4.0in}
\includegraphics{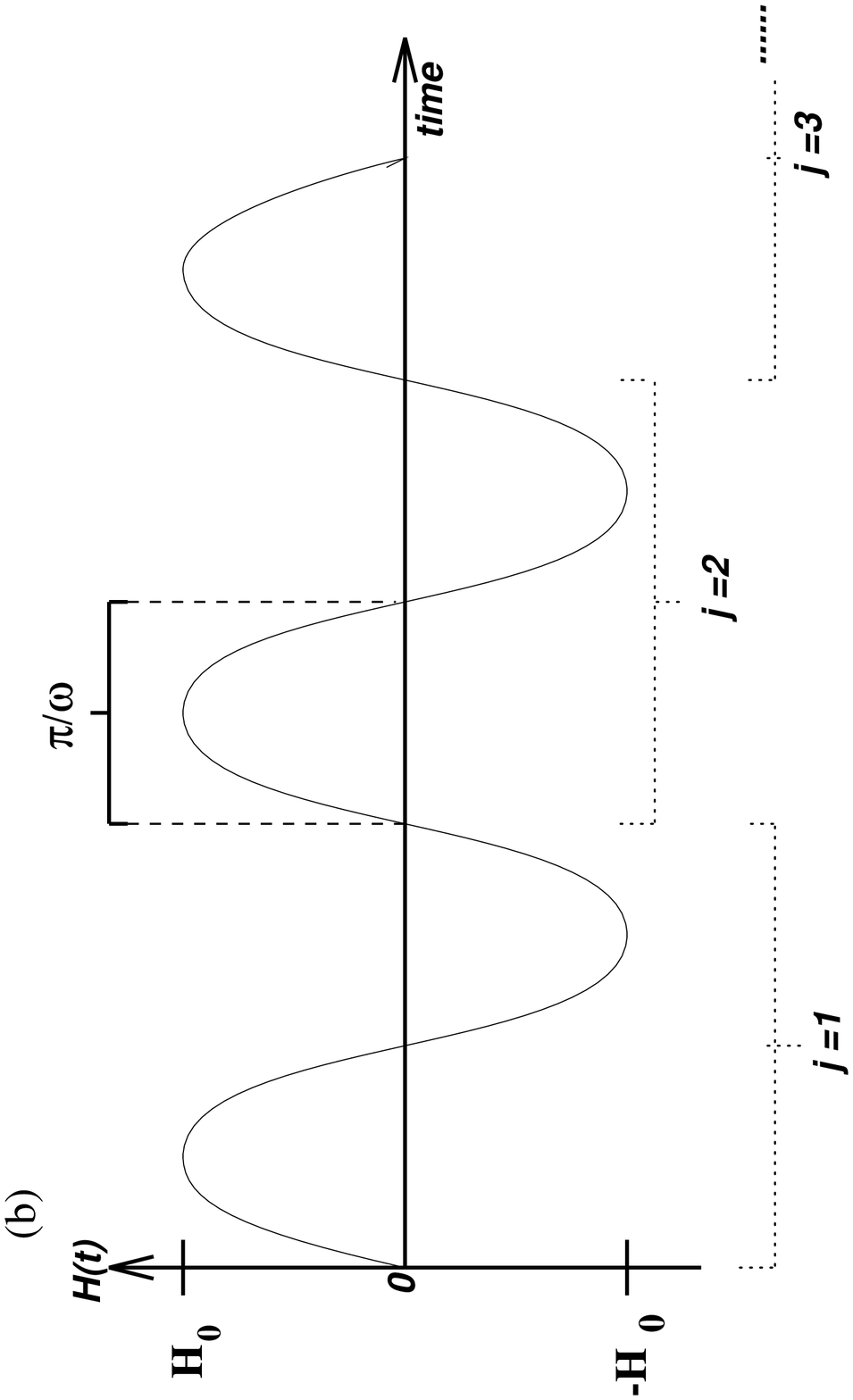}
\bigskip
\caption{\label{field_sch}
Schematic diagrams for calculation of the RTDs.}
\end{figure}

\end{document}